\documentclass[pre,aps,amsmath,twocolumn,noeprint]{revtex4-2}
\usepackage[T1]{fontenc}
\usepackage{graphicx}
\usepackage[utf8]{inputenc}
\usepackage[english]{babel}
\usepackage{lmodern}
\usepackage[hidelinks]{hyperref}
\usepackage[capitalize]{cleveref}
\usepackage{color}

\newcommand{\rhosr}{\chi_\star(\vec{r})}
\newcommand{\rhosidr}{\chi_\star^\mathrm{id}(\vec{r})}
\newcommand{\rhosc}{\chi_{\star, N}(\vec{r})}

\newcommand{\avg}[1]{\left\langle {#1} \right\rangle}
\newcommand{\rhor}[0]{\rho(\vec{r})}

\newcommand{\funcVr}[1]{\left. \frac{\delta #1}{\delta V_{\mathrm{ext}}(\vec{r})} \right |_{T,\mu} \!\!}

\renewcommand*{\vec}[1]{\mathbf{#1}}
\newcommand*{\Tr}[0]{\mathop{\mathrm{Tr}}\,}
\newcommand*\diff{\mathop{}\!\mathrm{d}}

\newcommand*{\intdN}[1]{\int\!\diff\vec{#1}^N\,}
\newcommand*{\intd}[1]{\int\!\diff\vec{#1}\,}
\newcommand*{\rhoop}[1]{\hat{\rho}(\vec{#1})}
\newcommand{\partd}[3]{\left. \frac{\partial #1}{\partial #2} \right |_{#3}}

\newcommand{\su}[1]{^\mathrm{#1}}
\newcommand{\ind}[1]{_\mathrm{#1}}

\newcommand{\cov}[1]{\,\mathrm{cov}\left( #1 \right)}
\newcommand{\covN}[1]{\,\mathrm{cov}_N\left( #1 \right)}

\renewcommand{\exp}[1]{\mathrm{exp} \left(#1\right)}

\newcommand{\wrt}{with respect to\ }
\newcommand{\V}{V\ind{ext}}
\newcommand{\Vr}{V\ind{ext}(\vec{r})}

\newcommand{\ctc}{\ensuremath{\chi_{T,N}(\vec{r})}}

\newcommand{\cmr}{\chi_\mu(\vec{r})}
\newcommand{\ctr}{\chi_T(\vec{r})}
\newcommand{\cmidr}{\chi_\mu^\mathrm{id}(\vec{r})}
\newcommand{\ctidr}{\chi_T^\mathrm{id}(\vec{r})}
\newcommand{\eqdef}{\equiv}
\newcommand{\phase}{\left(\vec{r}^N,\vec{p}^N\right)}
\newcommand{\eg}{e.g.\ }
\newcommand{\ie}{i.e.\ }

\newcommand{\Og}{\Omega [\rho]}
\newcommand{\Fid}{F\su{id} [\rho]}
\newcommand{\Fexc}{F\su{exc} [\rho]}

\newcommand{\contrib}{\thanks{Authors contributed equally to this work.}}

\begin{document}
\title{Local measures of fluctuations in inhomogeneous liquids: Statistical mechanics and illustrative applications}

\author{Tobias Eckert}
\contrib

\author{Nico C. X. Stuhlmüller}
\contrib

\author{Florian Sammüller}
\contrib

\author{Matthias Schmidt}
\email{Matthias.Schmidt@uni-bayreuth.de}
\affiliation{Theoretische Physik II, Physikalisches Institut,
  Universität Bayreuth, D-95440 Bayreuth, Germany}

\newcommand{\citeletter}{[{T.}~{Eckert}\ \emph {et~al.} \href {https://doi.org/10.1103/PhysRevLett.125.268004} {\bibfield  {journal} {\bibinfo  {journal} {Phys. Rev. Lett.}\ }\textbf {\bibinfo {volume} {125}},\ \bibinfo {pages} {268004} (\bibinfo {year} {2020})}]}

\begin{abstract}
  We show in detail how three one-body fluctuation profiles, namely the local compressibility, the local thermal susceptibility, and the reduced density, can be obtained from a statistical mechanical many-body description of classical particle-based systems.
  We present several different and equivalent routes to the definition of each fluctuation profile, facilitating their explicit numerical calculation in inhomogeneous equilibrium systems.
  This underlying framework is used for the derivation of further properties such as hard wall contact theorems and novel types of inhomogeneous one-body Ornstein-Zernike equations.
  The practical accessibility of all three fluctuation profiles is exemplified by grand canonical Monte Carlo simulations that we present for hard sphere, Gaussian core and Lennard-Jones fluids in confinement.
\end{abstract}

\maketitle


\section{Introduction}
The description of inhomogeneous soft matter is an important and challenging task both from a theoretical perspective as well as in practical applications arising throughout physics, chemistry and biology.
Many naturally occuring phenomena such as for instance water-repelling plant surfaces \cite{Koch2009}, the organization of cells and membranes \cite{Tanford1972} and protein interactions \cite{Ball2008} are attributed to be largely influenced by microscopic mechanical effects that occur upon immersion into a liquid, most commonly water.
Some of the underlying principles, e.g.\ that of hydrophobicity, are even deemed a necessity for the emergence of living matter \cite{Tanford1978}.
Likewise, these phenomena can be transferred to technological purposes, as was done e.g.\ in the design of self-cleaning materials \cite{Aytug2015,Ueda2013} or in the construction of specialized nanostructures \cite{Rasaiah2008}.

It is apparent that a bulk description is insufficient in situations where the behavior of a fluid is determined by an inhomogeneous external environment, e.g.\ due to being in contact with a substrate or by confinement in cavities or pores \cite{Berne2009,Chandler2005,Bocquet2010}.
Instead, to properly understand many of the above mentioned phenomena occuring in nature, but also to be able to tailor the behavior of soft matter for technological applications, a framework that accounts for microscopic spatial correlations imposed by an external potential landscape is needed.

From the viewpoint of statistical mechanics, molecular computer simulations \cite{Frenkel2023} as well as density functional theory (DFT) \cite{Evans1979,Hansen2013} are popular and powerful tools to reveal the structure of soft matter.
DFT is based on a minimization principle of the grand canonical potential $\Omega[\rho]$, which is expressed universally and in principle exactly as a functional of the spatially resolved one-body density profile $\rho(\vec{r})$.
Functional minimization of $\Omega[\rho]$ with respect to $\rho(\vec{r})$, carried out at fixed chemical potential $\mu$, temperature $T$, and external potential, then yields the physical equilibrium density profile.
Thus, $\rho(\vec{r})$ is alleviated from a mere observable to a central quantity that contains, via the connection to the grand potential $\Omega[\rho]$, all relevant thermodynamic information.

Given this situation, it is perhaps all the more surprising that, in recent years, not only the density profile $\rho(\vec{r})$ was considered in many of the fundamental phenomena mentioned above.
Instead, when solvophobic effects dominate the behavior of fluids, the local compressibility $\cmr = \partial \rho(\vec{r}) / \partial \mu$ has been shown to uncover much more information about structural correlations than the density profile alone \cite{Evans2015,Evans2015a,Evans2016,Evans2017,Archer2017,Evans2019,Remsing2019,Coe2022a,Coe2022,Coe2023}.
Solvophobicity, also called hydrophobicity if the solute is water, thereby refers to the tendency of a liquid phase to avoid contact with a substrate or solute.
This effect can occur over a broad range of length scales, from the solvation of single molecules to density depletion at macroscopic interfaces \cite{Chandler2005}.

Given the simple definition of the local compressibility as a parametric derivative of the density profile with respect to the chemical potential, it is clear that $\cmr$ is easily accessible in DFT calculations, and that it also can be obtained straightforwardly from molecular simulation.
Recently, further developments that go beyond the density profile as the primary means of characterizing inhomogeneous fluids have emerged.
Notable examples are the use of force density profiles within smart sampling techniques for molecular simulations \cite{Borgis2013,delasHeras2018,Rotenberg2020,Renner2023}, or reformulations of DFT to better account for two-body correlations \cite{Tschopp2020,Tschopp2021,Tschopp2022,Sammueller2023}.

Despite the demonstration of its practical relevance, the definition of $\cmr$ is somewhat ad hoc.
The question arises whether the observation of $\cmr$ being a better indicator of solvophobicity than $\rho(\vec{r})$ is merely relevant in specific situations, or whether it points to a more general theoretical framework.
Indeed, it has been shown by \citeauthor{Evans2015a} that the local compressibility is a suitable and rigorous quantification of localized fluctuations of the particle number \cite{Evans2015a}.
Thus, when these local fluctuations govern the behavior of inhomogeneous fluids -- as was shown to be the case for the multitude of processes related to hydrophobicity \cite{Jamadagni2011} -- the local compressibility enables a thorough study of the resulting physical effects.

Based on the above findings and the apparent importance of $\cmr$, one could still ask whether there are different types of density fluctuations that may be captured by additional one-body fluctuation fields, and whether a more fundamental mathematical framework than the mere chemical potential derivative of the density profile can be found.
In previous work, we proposed to complement the local compressibility by two additional observables such that a full set of fluctuation profiles is obtained, which reveals even more insight into the nature of local fluctuations \cite{Eckert2020}.
For the grand canonical ensemble, those additional one-body profiles are the local thermal susceptibility $\ctr$ and the reduced density $\rhosr$.

\citeauthor{Coe2022a} \cite{Coe2022a} have investigated the properties of these fluctuation profiles for critical drying based on DFT and on a mesoscopic treatment.
The authors concluded that for state points in the critical drying regime, all three measures, when scaled by their respective bulk values, show very similar behavior and the ratio $\chi_T(z) / \chi_\mu(z)$ is constant when the distance $z$ from the substrate is chosen to be in the gas-liquid interfacial region.
Such behavior can be expected, based on general scaling arguments of surface thermodynamics \cite{Coe2022a}.
Furthermore, addressing the microscopic behavior of electrolytes, \citeauthor{Cats2021} \cite{Cats2021} argue that the differential capacitance is a probe for the electric double layer structure and the bulk composition of an electrolyte.
Inspired by Ref.\ \cite{Eckert2020}, they formulated an expression for the differential capacitance by correlating the local density with the global charge in the system.
Given the significance of the fluctuation profiles in these previous works, we return to fundamentals and show in detail how the three fluctuation profiles can be obtained systematically and consistently from a many-body description of inhomogeneous equilibrium systems.

This paper is organized as follows.
Starting from statistical mechanics, as laid out in \cref{sec:grand-ensemble}, we derive functional generator expressions in \cref{sec:funct-deriv}, which give fluctuation measures as response functions to variation of the external potential $V_\mathrm{ext}(\vec{r})$.
Via a rewriting of derivatives, these expressions can be reformulated as in practice more easily accessible parametric derivatives with respect to the thermodynamic variables.
In \cref{sec:correlator_expressions}, we demonstrate that microscopic covariance forms exist for all three fluctuation profiles, which constitutes a further method for their numerical computation.
In \cref{sec:OZ}, one-body Orstein-Zernike equations for $\cmr$ and $\ctr$ are derived.
The fluctuation profiles are described for the hard sphere model in \cref{sec:hard_core} and hard wall contact theorems are given in \cref{sec:contact_values}.
While our derivations are formulated in a grand canonical setting, we explicitly point out differences that occur when using the canonical ensemble in \cref{sec:canonical}.

Based on grand canonical Monte Carlo (GCMC) simulations as described in \cref{sec:simulation}, we consider several standard situations using different interparticle interaction potentials, including hard spheres in \cref{sec:hard}, and Lennard-Jones particles and the Gaussian core model in \cref{sec:LJ_Gauss_confinement}.
The behaviour of the fluctuation profiles is studied primarily near substrates, where we observe markedly different behavior depending on both the particle and substrate type.
The effects of changing the thermodynamic state point are considered, which reveals characteristic behaviour of the fluctuation profiles as a means to probe and predict emerging phase transitions.
Lastly, we demonstrate the numerical calculation of fluctuation profiles for different ensembles in \cref{sec:cgc} and conclude in \cref{sec:conclusion}.


\section{Statistical mechanics of fluctuation profiles}
\label{sec:theory}

\subsection{Grand ensemble}
\label{sec:grand-ensemble}
We base our derivations of identities and properties of fluctuation profiles on the grand canonical ensemble (cf.\ \cref{sec:canonical} for a generalization to the canonical ensemble).
In the following, first our notation is introduced.

We consider systems of $N$ particles described by the Hamiltonian
\begin{equation}
  \label{eq:hamiltonian}
  H \phase =  \sum_{i=1}^N \frac{\vec{p}_i^2}{2m} + u(\vec{r}^{N})  + \sum_{i=1}^N V\ind{ext}(\vec{r}_i),
\end{equation}
with mass $m$, particle coordinates $\vec{r}^N \eqdef \{\vec{r}_{1}, \dots, \vec{r}_{N}\}$, and momenta $\vec{p}^N \eqdef \{\vec{p}_{1}, \dots, \vec{p}_{N}\}$ in $D$ spatial dimensions.
The particles interact both via an interparticle potential $u(\vec{r}^{N})$ and with an external potential $\Vr$, where $\vec{r}$ is a generic position variable.
In equilibrium, the grand canonical probability distribution function attains the Boltzmann form
\begin{align}
  \label{eq:prob-distribution}
  \Psi \phase  = \frac{\mathrm{e}^{ -\beta (H - \mu N) }}{\Xi}.
\end{align}
Here, $\Xi$ denotes the grand partition sum given by $\Xi = \Tr \mathrm{e}^{ -\beta (H - \mu N) }$ where $\beta = 1/(k_BT)$ with Boltzmann's constant $k_B$, temperature $T$ and chemical potential $\mu$.
The grand canonical trace is defined as
\begin{equation}
  \label{eq:trace}
  \Tr \cdot = \sum_{N=0}^{\infty} \frac{1}{N! h^{DN}} \intdN{r}\!\intdN{p} \cdot
\end{equation}
and it represents the sum of all particle numbers $N$ and phase space integral over all particle positions and momenta; $h$ denotes Planck's constant.
The grand canonical average of a phase space function $A \left( \vec{r}^N, \vec{p}^N \right)$ is defined as
\begin{equation}
  \label{eq:average}
  \avg{A} = \Tr A \Psi.
\end{equation}

The appropriate thermodynamic potential is the grand potential $\Omega = -k_BT \ln \Xi$, which is related to the internal energy $U$, entropy $S$ and average number of particles $\avg{N}$ via
\begin{equation}
  \label{eq:OmegaU}
  \Omega = U - TS - \mu \avg{N}.
\end{equation}
Parametric differentiation of $\Omega$ \wrt its natural thermodynamic variables yields
\begin{equation}
  \label{eq:Omega-diff}
  \partd{\Omega}{T}{\mu} = -S, \qquad
  \partd{\Omega}{\mu}{T} = -\avg{N}.
\end{equation}
Note that due to \cref{eq:Omega-diff}, the decomposition \eqref{eq:OmegaU} of the grand potential possesses the structure of a Legendre transform.
The state of a grand canonical system is specified by the values of the variables $T$, $\mu$ and $V$.
Since we consider the presence of a general external potential, we can express any finite system by including hard walls in $\Vr$.
Thus, we keep the volume $V$ fixed, without loss of generality, and omit the variable $V$ in the following.

\subsection{Functional and parametric derivatives}
\label{sec:funct-deriv}
We first show that one-body fluctuation profiles can be generated systematically as functional derivatives of appropriate global objects with respect to the external potential $\Vr$.
Additionally, it is demonstrated that the resulting fields can be obtained equivalently from parametric derivatives of the density profile with respect to thermodynamic variables.

We first recall the well-known relation \cite{Evans1979}
\begin{equation}
  \label{eq:Omega-rho}
  \funcVr{\Omega} = \rhor,
\end{equation}
which expresses that the density profile $\rhor \eqdef \avg{\sum_{i=1}^N \delta(\vec{r}_i-\vec{r})}$ is generated as a response function of the grand potential $\Omega$ upon changing $\Vr$.
Using the decomposition \eqref{eq:OmegaU} of the grand potential into $U$, $S$ and $\avg{N}$, we take the functional derivative on the left hand side of \cref{eq:Omega-rho} separately for each term,
\begin{equation}
\label{eq:Omega-func}
  \rhor = \funcVr{U} - T \funcVr{S} - \mu \funcVr{\avg{N}}.
\end{equation}
As before we have made explicit in the notation that the functional derivative is evaluated at constant thermodynamic state point $T, \mu$.
\Cref{eq:Omega-func} is a fundamental relation that we take as motivation to naturally consider the three terms on the right hand side separately.

We first focus on the last two terms on the right hand side of \cref{eq:Omega-func} in order to reformulate the functional derivatives of $S$ and $\langle N \rangle$ as parametric derivatives of the density profile.
In the latter case, this is achieved by recalling $\langle N \rangle = - \partial \Omega / \partial \mu$ from relation \eqref{eq:Omega-diff}, which is inserted into the functional derivative $\delta \langle N \rangle / \delta V_\mathrm{ext}(\vec{r})$.
Then, we perform an exchange of functional and parametric derivative, thereby paying attention to the variables that are kept fixed.
After this exchange, the density profile can be identified due to \cref{eq:Omega-rho} and one obtains the identity
\begin{equation}
  \label{eq:func-partial-rho-N}
  \begin{split}
  \funcVr{\avg{N}} &= -\funcVr{} \partd{\Omega}{\mu}{T} \\
                   &= -\partd{}{\mu}{T} \funcVr{\Omega} = -\partd{\rhor}{\mu}{T}.
  \end{split}
\end{equation}
Here, the external potential is kept fixed when evaluating the parametric derivative.
We can identify from \cref{eq:func-partial-rho-N} the local compressibility
\begin{align}
  \label{eq:chi-mu-func}
  \cmr &\eqdef -\funcVr{\avg{N}} \\
  \label{eq:chi-mu-para}
       &= \partd{\rhor}{\mu}{T},
\end{align}
as a first fluctuation profile, which has been introduced by \citeauthor{Evans2015} \cite{Evans2015} in the parametric derivative form \eqref{eq:chi-mu-para}.
The above derivation reveals that $\cmr$ can be defined equivalently as a functional generator expression of the average number of particles $\langle N \rangle$ with respect to changes in the external potential \cite{Evans2015a,Eckert2020}.

For the functional derivative of the entropic term in \cref{eq:Omega-func}, analogous reasoning as in \cref{eq:func-partial-rho-N} leads to
\begin{equation}
  \label{eq:func-partial-rho-S}
  \begin{split}
  \funcVr{S} &= -\funcVr{} \partd{\Omega}{T}{\mu} \\
             &=-\partd{}{T}{\mu} \funcVr{\Omega} = -\partd{\rhor}{T}{\mu}
  \end{split}
\end{equation}
Therefore, a second fluctuation profile emerges from \cref{eq:func-partial-rho-S}, which we define \cite{Eckert2020} as the local thermal susceptibility
\begin{align}
  \label{eq:chi-T-func}
  \ctr &\eqdef -\funcVr{S} \\
  \label{eq:chi-T-para}
       &= \partd{\rhor}{T}{\mu}.
\end{align}
Formally, $\ctr$ possesses an analogous structure to $\cmr$, as it can be acquired either from a functional generator expression with respect to variation of $\Vr$, or alternatively as a parametric derivative of $\rhor$.

We define the remaining first term of the right hand side of \cref{eq:Omega-func} as the reduced density profile:
\begin{equation}
  \label{eq:rhos-generator}
  \rhosr \eqdef \funcVr{U}.
\end{equation}
In contrast to the availability of $S$ or $\langle N \rangle$ as a derivative of the grand potential, cf.\ \cref{eq:Omega-diff}, the internal energy $U$ possesses no such form.
Therefore, also no reformulation into a single parametric derivative of the density profile exists for $\rhosr$.
However, by rearranging \cref{eq:Omega-func} and inserting the functional generator definitions in \cref{eq:chi-mu-func,eq:chi-T-func,eq:rhos-generator}, we can express the reduced density as
\begin{equation}
  \label{eq:rhos}
  \rhosr = \rhor - T \ctr - \mu \cmr.
\end{equation}
We recall that $\cmr$ and $\ctr$ can be formulated equivalently as parametric derivatives of $\rhor$, cf.\ \cref{eq:chi-mu-para,eq:chi-T-para}; thus the reduced density profile can be obtained without the need of explicitly evaluating the functional derivative \eqref{eq:rhos-generator}.
Additionally, \cref{eq:rhos} possesses for fixed position $\vec{r}$ the structure of a local Legendre transform
\begin{equation}
  \rhosr = \rhor - T \partd{\rhor}{T}{\mu} - \mu \partd{\rhor}{\mu}{T}.
\end{equation}
This is analogous to the global Legendre structure of the grand potential
\begin{equation}
  U = \Omega - T \partd{\Omega}{T}{\mu} - \mu \partd{\Omega}{\mu}{T},
\end{equation}
resulting from inserting \cref{eq:Omega-diff} into \cref{eq:OmegaU}.

As a summary, starting from \cref{eq:Omega-func}, we have obtained the local compressibility $\cmr$, the local thermal susceptibility $\ctr$ and the reduced density $\rhosr$, which constitute a set of fluctuation profiles in the grand canonical ensemble.
The local compressibility $\cmr$ as given via its parametric derivative form in \cref{eq:chi-mu-para} has been studied extensively in Refs.\ \cite{Evans2015a,Evans2016,Evans2017,Archer2017,Evans2019,Remsing2019}.
The authors of these works conclude that $\cmr$ is a better measure of drying than is $\rhor$.
In the case of drying, the local density of a liquid adsorbed against a solvophobic wall falls below the fluid bulk density even all the way down to the vapour density.
This vapour layer in front of the wall induces particularly large localized fluctuations in the number of particles, which are captured by $\cmr$ due to \cref{eq:chi-mu-para}.
The above derivation shows that additional fluctuation profiles $\ctr$ and $\rhosr$ exist, which correspond to entropic and energetic fluctuations, cf.\ \cref{eq:chi-T-func} and \cref{eq:rhos-generator} respectively.
By providing means to capture those fluctuations, the local thermal susceptibility $\ctr$ and the reduced density $\rhosr$ therefore complement $\cmr$ in a way that promises to aid the analysis of fluctuations in inhomogeneous fluids, see Ref.\ \cite{Coe2022a,Coe2022} for an in-depth analysis for the case of critical drying.

For completeness, the bulk counterparts of $\cmr$ and $\ctr$ are defined by replacing the one-body density $\rhor$ with the bulk density $\rho^b = \langle N \rangle / V$ in the parametric derivative definitions \eqref{eq:chi-mu-para} and \eqref{eq:chi-T-para}, \ie $\chi_\mu^b = \partial \rho^b / \partial \mu$ and $\chi_T^b = \partial \rho^b / \partial T$.
Using \cref{eq:Omega-diff},
\begin{equation}
  \label{eq:chi_mu_bulk_positive}
  \chi_\mu^b = \frac{1}{V} \left.\frac{\partial \langle N \rangle}{\partial \mu}\right|_{V, T} = - \frac{1}{V} \left.\frac{\partial^2 \Omega}{\partial \mu^2}\right|_{V, T} > 0,
\end{equation}
where the inequality follows due to $\Omega(\mu)$ being a concave function.
Hence, $\chi_\mu^b$ is always positive even when interparticle interactions are present.
In contrast to the bulk value of the chemical susceptibility being positive, this property does not hold locally for $\chi_\mu(\vec{r})$, as will be shown in the simulation results in \cref{sec:simulation}.

The bulk thermal susceptibility $\chi_T^b$ can be positive as well as negative since it is the off-diagonal element $\partial^2 \Omega / (\partial \mu \partial T)$ of the Hessian matrix of $\Omega$.
Hence, $\ctr$, the localized counterpart to $\chi_T^b$, can also take on positive as well as negative values.

\subsection{Correlator expressions}
\label{sec:correlator_expressions}
Expressing $\cmr$ and $\ctr$ via \cref{eq:chi-mu-para,eq:chi-T-para} as parametric derivatives of $\rhor$ already provides an intuitive interpretation of these functions as a local response of the density profile to changes in $\mu$ and $T$, respectively.
Unlike the functional generator expressions, parametric derivatives are easily accessible in simulations, albeit still needing multiple runs carried out at neighboring state points with different values of $\mu$ or $T$ or relying on histogram methods.
Especially near phase coexistence, small shifts of the thermodynamic state point can induce large responses in $\rhor$ \cite{Evans2015,Evans2015a}, which makes the finite differences difficult to evaluate accurately.
In the following, an alternative method for the derivation of fluctuation profiles is presented, which avoids the evaluation of parametric derivatives and reveals the accessibility of $\cmr$, $\ctr$ and $\rhosr$ at fixed thermodynamic parameters $\mu$ and $T$.

The basis of the reasoning is the analytical evaluation of the functional derivatives in \cref{eq:Omega-func}.
This is possible in equilibrium, as the grand canonical probability distribution function $\Psi$ is known for a system described by a Hamiltonian $H$, cf.\ the familiar Boltzmann form \eqref{eq:prob-distribution} of $\Psi$.
We show in the following that functional differentiation then yields exact microscopic expressions for $\cmr$ \cite{Evans2015a}, $\ctr$ and $\rhosr$, which contain covariances of the density operator $\hat{\rho}(\vec{r}) = \sum_{i=1}^N \delta(\vec{r}_i-\vec{r})$ with $N$ and $H$.
The covariance of two phase space functions $A \left( \vec{r}^N, \vec{p}^N \right)$ and $B \left( \vec{r}^N, \vec{p}^N \right)$ is thereby defined as
\begin{equation}
  \label{eq:cov}
  \cov{A,B} \eqdef \avg{AB} - \avg{A}\avg{B},
\end{equation}
where we recall the definition of the grand ensemble average \eqref{eq:average} as indicated by the angular brackets.

To evaluate the functional generator expressions given in \cref{eq:chi-mu-func,eq:chi-T-func,eq:rhos-generator}, we need to differentiate $\avg{N}$, $S$ and $U$ functionally \wrt $\Vr$.
Thereby, functional derivatives of $H$, $\Xi$ and $\Psi$ are required in turn.
We present these in detail in \cref{sec:func-derivatives}.

We start with the response of the average number of particles $\avg{N} = \Tr N \Psi$ with respect to changes in the external potential according to \cref{eq:chi-mu-func} and obtain
\begin{align}
  \label{eq:func-N-2}
  \funcVr{\avg{N}} &= \Tr N \funcVr{\Psi} \\
  \label{eq:func-N-1}
                   &= -\beta \Tr N \left[\rhoop{r} - \rhor \right] \Psi \\
                   &= -\beta \left[ \avg{N\rhoop{r}} - \avg{N} \rhor \right] \\
  \label{eq:func-N}
                   &= -\beta \cov{N,\rhoop{r}}.
\end{align}
For the derivation of $\delta \Psi / \delta V_\mathrm{ext}(\vec{r}) = -\beta \left[ \rhoop{r} - \rhor \right] \Psi$, see \cref{eq:func-psi} in \cref{sec:func-derivatives}.
The density profile $\rhor$ in \cref{eq:func-N-1} can be taken out of the average, which yields \cref{eq:func-N} after identification of the covariance.
As the left hand side of \cref{eq:func-N-2} is the negative functional generator expression \eqref{eq:chi-mu-func} of the local compressibility $\cmr$ we obtain the identity \cite{Evans2015a,Evans2017}
\begin{equation}
  \label{eq:chi_mu_cov}
  \cmr = \beta \cov{N,\rhoop{r}}.
\end{equation}
It is striking that the total number of particles $N$ appears in the covariance, despite $\cmr$ being a localized (one-body) quantity.
From inspecting \cref{eq:func-N}, we observe a general mechanism: if an operator $A$ is not explicitly dependent on $\Vr$, e.g.\ $A = N$, then the functional derivative $\delta \langle A \rangle / \delta \Vr |_{T, \mu}$ yields $-\beta \cov{A, \rhoop{r}}$.

We turn to the first term of the right hand side of \cref{eq:Omega-func} and hence consider the response of $U = \avg{H} = \Tr H \Psi$ to changes in $\Vr$.
Clearly, the Hamiltonian \eqref{eq:hamiltonian} has an explicit dependence on $\Vr$.
Therefore, unlike in \cref{eq:chi_mu_cov}, additional terms besides the covariance of $H$ and $- \beta \rhoop{r}$ are expected to occur.
Using the product rule of differentiation, we evaluate
\begin{align}
  \funcVr{\avg{H}} &= \Tr \funcVr{H} \Psi + \Tr H  \funcVr{\Psi} \\
  \label{eq:func-avgH-1}
         &= \Tr \rhoop{r} \Psi  -\beta \Tr H \left[ \rhoop{r} - \rhor \right] \Psi \\
  \label{eq:func-avgH}
         &= \rhor  - \beta \cov{H,\rhoop{r}}.
\end{align}
For the derivation of $\delta H / \delta \Vr = \rhoop{r}$, we refer the reader to \cref{eq:func-H} in \cref{sec:func-derivatives}.
Besides the energy-density covariance, an additional $\rhor$ contribution emerges in \cref{eq:func-avgH}, which is the ``direct response'' of $H$ upon changes in $\Vr$.
We recall the definition \eqref{eq:rhos-generator} of $\rhosr$ and find from \cref{eq:func-avgH} the covariance form of the reduced density profile
\begin{align}
  \label{eq:rhos-cov}
  \rhosr = \rhor - \beta \cov{H,\rhoop{r}}.
\end{align}

We next evaluate the response of the entropy $S$ \wrt changes in $\Vr$ to derive the covariance form of the local thermal susceptibility $\ctr$.
The entropy operator is defined \cite{Jaynes1965} as
\begin{equation}
  \label{eq:S-hat}
  \hat{S} = -k_B \ln\Psi,
\end{equation}
and the total entropy then follows from $S = \langle \hat{S} \rangle = -k_B \Tr \Psi \ln\Psi$.
In contrast to the particle number or the Hamiltonian, $\hat{S}$ depends explicitly on the many-body distribution function $\Psi$ as given in \cref{eq:prob-distribution}.

Functional differentiation of $S$ with respect to $\Vr$ then leads to
\begin{align}
  \funcVr{\langle \hat{S} \rangle} &= -k_B \Tr \funcVr{ \Psi\ln\Psi} \\
  \label{eq:func-S-3}
             &= -k_B \Tr ( 1 + \ln\Psi )\funcVr{\Psi} \\
  \label{eq:func-S-2}
             &= \beta k_B\Tr ( 1 + \ln\Psi ) (\rhoop{r} - \rhor ) \Psi \\
  \label{eq:func-S-1}
             &= -\beta \Tr \left( \hat{S} \rhoop{r} - S \rhor \right) \Psi \\
  \label{eq:func-S}
             &= -\beta \cov{\hat{S},\rhoop{r}}.
\end{align}
The product rule has been used to obtain \cref{eq:func-S-3} and the functional derivative $\delta \Psi / \delta V_\mathrm{ext}(\vec{r}) = -\beta \left[ \rhoop{r} - \rhor \right] \Psi$ has been evaluated in \cref{eq:func-psi}.
This yields \cref{eq:func-S-2}, where one can successively identify the entropy operator \eqref{eq:S-hat} and the covariance \eqref{eq:cov} to obtain \cref{eq:func-S}.
Comparison with the functional generator expression \eqref{eq:chi-T-func} for $\ctr$ establishes
\begin{equation}
  \label{eq:chi-T-covS}
  \ctr = \beta \cov{\hat{S},\rhoop{r}}.
\end{equation}
Due to \cref{eq:chi-T-covS}, $\ctr$ can be interpreted as a residual entropy, and it is related to the entropy densities introduced in Refs.\ \cite{Wallace1987,Schmidt2011}.

While \cref{eq:chi-T-covS} constitutes a successful rewriting of $\ctr$ into a covariance form, its practical use is not fully revealed yet.
This is due to the fact that $\hat{S}$ cannot be obtained straightforwardly from molecular simulations as it depends explicity on the partition function, cf.\ \cref{eq:S-hat}.
However, one can express \cref{eq:chi-T-covS} via more easily accessible quantities as follows.

We insert the definition \eqref{eq:S-hat} of the entropy operator into \cref{eq:chi-T-covS} which yields
\begin{align}
  \label{eq:cov-S-2}
  \cov{\hat{S},\rhoop{r}} &= -k_B \cov{\ln \Psi,\rhoop{r}} \\
  \label{eq:cov-S-1}
                          &= k_B \cov{\beta(H-\mu N) + \ln\Xi,\rhoop{r}}\\
  \label{eq:cov-S-0}
                          &= \frac{1}{T} \cov{H,\rhoop{r}} - \frac{\mu}{T} \cov{N,\rhoop{r}},
\end{align}
where the Boltzmann form \eqref{eq:prob-distribution} of $\Psi$ has been used in \cref{eq:cov-S-2}.
As $\ln\Xi = -\beta \Omega$ is a constant with respect to the phase space variables, its covariance with $\rhoop{r}$ has vanished in \cref{eq:cov-S-0}, and one arrives at
\begin{equation}
  \label{eq:chi-T-cov}
  \ctr = \frac{1}{k_BT^2} \cov{H,\rhoop{r}} - \frac{\mu}{k_BT^2} \cov{N,\rhoop{r}}.
\end{equation}
\Cref{eq:chi-T-cov} only contains covariances of the density operator with $H$ and $N$ respectively.
Contrary to \cref{eq:chi-T-covS}, these covariances can be sampled in a standard manner in GCMC simulations without the need for methods such as thermodynamic integration \cite{Mitchell1991,Frenkel2023}.

Hence, practically useful covariance forms of $\cmr$, $\ctr$ and $\rhosr$ as given in \cref{eq:chi_mu_cov,eq:rhos-cov,eq:chi-T-cov} enable the sampling of all three fluctuation profiles directly within a GCMC simulation at fixed thermodynamic state point.
This method is applied in \cref{sec:simulation}, where the fluctuation profiles of prototypical fluid models in confinement are investigated within GCMC simulations.

\subsection{Fluctuation Ornstein-Zernike relations}
\label{sec:OZ}
The standard inhomogeneous two-body Ornstein-Zernike (OZ) relation is an integral equation that connects the pair correlation function $h(\vec{r}_1, \vec{r}_2)$ with the direct correlation function $c_2(\vec{r}_1, \vec{r}_2)$ and the density profile as follows \cite{Ornstein1914,Hansen2013} (for a pedagogical introduction, see \eg Ref.\ \cite{Schmidt2022})
\begin{equation}
  \label{eq:OZ_standard}
  h(\vec{r}_{1}, \vec{r}_{2}) = c_2(\vec{r}_{1}, \vec{r}_{2}) + \intd{r'} \rho(\vec{r'}) c_2(\vec{r}_1, \vec{r'}) h(\vec{r'}, \vec{r}_{2}).
\end{equation}
We show that analogous integral equations can be derived for the fluctuation profiles $\cmr$ and $\ctr$ and that these relations are of simpler, one-body nature.
Due to their resemblance to the original OZ equation \eqref{eq:OZ_standard}, we will also refer to the resulting integral equations for $\cmr$ and $\ctr$ as fluctuation Ornstein-Zernike equations.

We proceed similar to the concepts of Refs.\ \cite{Brader2013,Brader2014} and start with the fundamental minimization principle of the grand potential $\Omega[\rho]$, which is expressed as a functional of a (trial) density profile $\rhor$ that attains its minimum at the physical equilibrium density profile,
\begin{equation}
  \label{eq:omega-EL}
  \frac{\delta \Og}{\delta \rho(\vec{r})} = 0, \qquad \text{(min)}.
\end{equation}
The grand potential is composed of ideal and excess intrinsic free energy functionals, $\Fid$ and $\Fexc$ respectively, as well as of external contributions \cite{Evans1979},
\begin{equation}
  \label{eq:omega-DFT-splitting}
  \Omega[\rho] = \Fid + \Fexc + \intd{r'} (V_\mathrm{ext}(\vec{r'}) - \mu) \rho(\vec{r'}).
\end{equation}

Explicit evaluation of the functional derivative in \eqref{eq:omega-EL} after insertion of \cref{eq:omega-DFT-splitting} yields (upon exchanging the left and right hand sides)
\begin{equation}
  \label{eq:grand-potential-DFT}
  0 = k_B T \ln(\Lambda^D \rho(\vec{r})) + \frac{\delta \Fexc}{\delta \rho(\vec{r})} + V_\mathrm{ext}(\vec{r}) - \mu,
\end{equation}
which is the standard Euler-Lagrange equation of DFT.
We have used the exact expression for the ideal gas free energy functional $\Fid = k_BT\intd{r} \rhor \left(\ln(\Lambda^D \rhor) - 1\right)$ with the thermal wavelength $\Lambda = h / \sqrt{2 \pi m k_B T}$ and can identify the second term on the right hand side of \cref{eq:grand-potential-DFT} as a contribution proportional to the one-body direct correlation function $c_1(\vec{r}) = - \beta \delta \Fexc / \delta\rho(\vec{r})$ \cite{Hansen2013}.

We now differentiate both sides of \cref{eq:grand-potential-DFT} \wrt $\mu$ upon keeping $T$ and $\Vr$ fixed.
This amounts to the physical situation of isothermally changing $\mu$ in the given system,
\begin{align}
  \label{eq:OZ chi_mu}
  0 &= \partd{}{\mu}{T} \left( k_B T \ln(\Lambda^D \rho(\vec{r})) + \frac{\delta \Fexc}{\delta \rho(\vec{r})} + V_\mathrm{ext}(\vec{r}) - \mu \right) \\
  \label{eq:OZ chi_mu+1}
    &= \frac{k_B T}{\rho(\vec{r})} \partd{\rhor}{\mu}{T} + \intd{r'} \frac{\delta^2 \Fexc}{\delta \rho(\vec{r}) \delta \rho(\vec{r'})} \partd{\rho(\vec{r'})}{\mu}{T}  - 1 \\
  \label{eq:OZ chi_mu+2}
    &= \frac{k_B T}{\rho(\vec{r})} \chi_\mu(\vec{r}) - k_B T \intd{r'} c_2(\vec{r}, \vec{r'}) \chi_\mu(\vec{r'}) - 1.
\end{align}
The functional chain rule has been applied to obtain the second term in \cref{eq:OZ chi_mu+1} and $\chi_\mu(\vec{r})$ is identified in \cref{eq:OZ chi_mu+2} via its parametric derivative definition \eqref{eq:chi-mu-para}.
Furthermore, $c_2(\vec{r}, \vec{r'}) = -\beta \delta^2 F^\mathrm{exc}/\delta \rho(\vec{r}) \delta \rho(\vec{r'})$ denotes the two-body direct correlation function \cite{Hansen2013}.

Multiplying \cref{eq:OZ chi_mu+2} by $\beta\rhor$ then yields a one-body Ornstein-Zernike relation for the local compressibility
\begin{equation}
  \label{eq:OZ chi_mu final}
  \chi_\mu(\vec{r}) = \rho(\vec{r}) \intd{r'} c_2(\vec{r}, \vec{r'}) \chi_\mu(\vec{r'}) + \beta \rhor.
\end{equation}

To obtain a similar equation for $\ctr$, the parametric derivative of the Euler-Lagrange \cref{eq:grand-potential-DFT} \wrt $T$ is calculated, which yields
\begin{align}
  \label{eq:OZ chi_T+1}
  0 &= \partd{}{T}{\mu} \left( k_B T \ln(\Lambda^D \rho(\vec{r})) + \frac{\delta \Fexc}{\delta \rho(\vec{r})} + V_\mathrm{ext}(\vec{r}) - \mu \right) \\
  \label{eq:OZ chi_T}
  \begin{split}
    &= k_B \left( \ln(\Lambda^D \rhor) + T \frac{\chi_T(\vec{r})}{\rhor} - \frac{D}{2} \right)\\
    &\quad - k_B T \left( \frac{\partial c_1(\vec{r})}{\partial T} + \frac{c_1(\vec{r})}{T} + \intd{r'} c_2(\vec{r}, \vec{r'}) \chi_T(\vec{r'}) \right).
  \end{split}
\end{align}
The derivatives of $\Vr$ and $\mu$ \wrt $T$ vanish, since neither $\Vr$ nor $\mu$ depend on $T$.
The dependence of $\delta \Fexc / \delta \rho(\vec{r}) = - k_B T c_1(\vec{r})$ on $T$ results in the second line of \cref{eq:OZ chi_T}.
The spatial integral over $c_2(\vec{r}, \vec{r}')$ and $\chi_T(\vec{r}')$ emerges again from the functional chain rule similar to \cref{eq:OZ chi_mu+1}.
Additionally, the thermal wavelength $\Lambda$ contains an explicit dependence on $T$, such that
\begin{equation}
  \label{eq:thermal_wavelength_derivative}
T \partd{\Lambda^D}{T}{\mu} = -\frac{D}{2} \Lambda^D,
\end{equation}
which is considered in the differentiation of $\ln(\Lambda^D \rhor)$ in \cref{eq:OZ chi_T+1} to obtain the kinetic contributions (proportional to $D/2$) to $\ctr$.
For the convention of the thermal wavelength set to the particle diameter $\Lambda = \sigma$, all of the following equations still apply, but with $D$ set to zero.
Taking the temperature dependence of $\Lambda$ into account explicitly also leads to a transformation of the chemical potential $\mu$, which is further illustrated in Appendix \ref{sec:lambda}.

Multiplying both sides of \cref{eq:OZ chi_T} with $\beta\rhor$ and rearranging terms results in an Ornstein-Zernike relation for the local thermal susceptibility
\begin{equation}
  \label{eq:OZ chi_T final}
  \begin{split}
    \chi_T(\vec{r}) = \rho(\vec{r}) &\left[ \frac{\partial c_1(\vec{r})}{\partial T} + \frac{c_1(\vec{r})}{T} + \intd{r'} c_2(\vec{r}, \vec{r'}) \chi_T(\vec{r'}) \right.\\
      &\left. + \frac{1}{T} \left( \frac{D}{2} - \ln(\Lambda^D \rho(\vec{r})) \right) \right].
\end{split}
\end{equation}

The fluctuation OZ \cref{eq:OZ chi_mu final,eq:OZ chi_T final} can be simplified further by performing a splitting of $\cmr$ and $\ctr$ into ideal (superscript $\mathrm{id}$) and excess (superscript $\mathrm{exc}$) contributions.
Details of an analysis of the local compressibility and local thermal susceptibility for the ideal gas are given in \cref{sec:ideal-gas}.
The results are
\begin{align}
  \label{eq:chi_mu_id}
  \chi_\mu^\mathrm{id} &= \beta \rhor,\\
  \label{eq:chi_T_id}
  \chi_T^\mathrm{id} &= \frac{1}{T} \left( \frac{D}{2} - \ln(\Lambda^D \rho(\vec{r})) \right) \rho(\vec{r}).
\end{align}
For an interacting system, the excess parts of $\cmr$ and $\ctr$ are defined by evaluating \cref{eq:chi_mu_id,eq:chi_T_id} at the equilibrium density profile $\rho(\vec{r})$ of the interacting system and considering the difference to the full fluctuation profiles, i.e.\
\begin{align}
  \chi_\mu^\mathrm{exc}(\vec{r}) &\eqdef \chi_\mu(\vec{r}) - \chi_\mu^\mathrm{id}(\vec{r}) = \chi_\mu(\vec{r}) - \beta\rho(\vec{r}), \\
  \label{eq:ct-id}
  \chi_T^\mathrm{exc}(\vec{r}) &\eqdef \chi_T(\vec{r}) - \chi_T^\mathrm{id}(\vec{r}) = \notag \\
                                 &= \chi_T(\vec{r}) - \frac{1}{T} \left( \frac{D}{2} - \ln(\Lambda^D \rho(\vec{r})) \right) \rho(\vec{r}).
\end{align}

One recognizes that this splitting appears naturally in the fluctuation OZ \cref{eq:OZ chi_mu final,eq:OZ chi_T final}, which can hence be rewritten succinctly as
\begin{align}
  \label{eq:OZ chi_mu final exc}
  \chi_\mu^\mathrm{exc}(\vec{r}) &= \rho(\vec{r}) \intd{r'} c_2(\vec{r}, \vec{r'}) \chi_\mu(\vec{r'}), \\
  \label{eq:OZ chi_T final exc}
    \chi_T^\mathrm{exc}(\vec{r}) &= \rho(\vec{r}) \left[ \frac{\partial c_1(\vec{r})}{\partial T} + \frac{c_1(\vec{r})}{T} + \intd{r'} c_2(\vec{r}, \vec{r'}) \chi_T(\vec{r'}) \right].
\end{align}

\subsection{Hard sphere fluctuation profiles}
\label{sec:hard_core}
For model fluids consisting of hard spheres, further simplifications arise.
We start with the fluctuation Ornstein-Zernike equation \eqref{eq:OZ chi_T final}, which simplifies for hard spheres to
\begin{equation}
  \chi_T^\mathrm{HS}(\vec{r}) = \rho(\vec{r}) \left( \frac{c_1(\vec{r})}{T} + \intd{r'} c_2(\vec{r}, \vec{r'}) \chi_T(\vec{r'}) \right),
\end{equation}
since $\partial c_1(\vec{r}) / \partial T = 0$ holds in this case.

Furthermore, as the internal interaction potential vanishes for all allowed microstates in the case of hard sphere interactions, i.e.\ $u(\vec{r}^N) = 0$ in \cref{eq:hamiltonian}, the correlator expression for the local compressibility \eqref{eq:chi-T-cov} reduces to

\begin{align}
  &T \chi_T(\vec{r}) \nonumber \\
  \label{eq:chi_T_hard_no_ext}
  &\quad= \cov{\beta \sum_{i = 1}^N \left( \frac{\vec{p}_i^2}{2 m} + V_\mathrm{ext}(\vec{r}_i) \right), \rhoop{\vec{r}}} - \mu \chi_\mu(\vec{r})\\
  &\quad= \left( \frac{D}{2} k_B T - \mu \right) \chi_\mu(\vec{r}) + \cov{\beta \sum_{i = 1}^N V_\mathrm{ext}(\vec{r}_i), \rhoop{\vec{r}}}\\
  \label{eq:chi_T_hard_density_cov}
  &\quad= \left( \frac{D}{2} k_B T - \mu \right) \chi_\mu(\vec{r}) + \beta \intd{r'} H_2(\vec{r}, \vec{r'}) V_\mathrm{ext}(\vec{r'}).
\end{align}

The equipartition theorem has been used for the evaluation of the kinetic energy part in \cref{eq:chi_T_hard_no_ext}.
Despite the local fluctuations being one-body fields, \cref{eq:chi_T_hard_density_cov} reveals explicitly that they also hold information about many-body correlation effects in interacting systems.
The two-body density-density covariance $H_2(\vec{r}, \vec{r}') = \cov{\rhoop{\vec{r}}, \rhoop{\vec{r}'}}$ mediates non-local effects of the external potential, which are hence also captured in $\ctr$.

For the case of a bulk system, \ie for vanishing external potential $V_\mathrm{ext}(\vec{r}) = 0$, or in systems with only hard walls, we find
\begin{equation}
  \label{eq:chi_T_chi_mu_hard}
  \frac{\chi_T(\vec{r})}{\chi_\mu(\vec{r})} = \frac{D}{2} k_B - \frac{\mu}{T},
\end{equation}
by rearrangement of \cref{eq:chi_T_hard_density_cov}, which makes the ratio of thermal and chemical fluctuations a constant, independent of position.

\Cref{eq:chi_T_chi_mu_hard} can be considered to be a contact theorem, as it (trivially) relates the contact value of fluctuation profiles at the hard wall to bulk quantities of the fluid.
In the following section, a more general contact theorem for $\ctr$ is derived that holds for arbitrary particle interactions, and which reduces to \cref{eq:chi_T_chi_mu_hard} in the case of hard spheres.

\subsection{Fluctuation hard wall contact theorems}
\label{sec:contact_values}
\citeauthor{Evans2015} \cite{Evans2015} derived the following contact theorem for the local compressibility,
\begin{equation}
  \label{eq:cmucontact}
  \chi_\mu(0^+) = \beta \rho(\infty),
\end{equation}
for a planar geometry with distance $x$ from the hard wall and $x \rightarrow \infty$ for the bulk fluid.
Following their derivation, we show that a contact theorem can be formulated for the local thermal susceptibility, which establishes connections to the internal energy of the system.

To derive the contact value of $\ctr$, we differentiate the well-known contact theorem of the density \cite{Tschopp2022},
\begin{equation}
  \label{eq:2}
  \rho(0^+) = \beta p(\mu,T),
\end{equation}
with respect to temperature, which yields
\begin{equation}
  \label{eq:6}
  \chi_T(0^+) = \beta \left( \frac{\partial p(\mu,T)}{\partial T} - \frac{p(\mu,T)}{T} \right).
\end{equation}
Here $p(\mu,T)$ indicates the bulk pressure.
Substituting the Gibbs-Duhem equation
\begin{equation}
  \label{eq:gibbsduhem}
  0 = S\diff T - V\diff p + \langle N \rangle \diff\mu,
\end{equation}
in the first term on the right hand side of \cref{eq:6} and $\Omega = - pV$ into the second term results in
\begin{align}
  \chi_T(0^+) &= \frac{ TS + \Omega}{Vk_BT^2} \\
              &= \frac{\langle H \rangle - \mu \langle N \rangle}{Vk_BT^2},
\end{align}
from which we obtain
\begin{equation}
  \label{eq:ctcontact}
  \chi_T(0^+) = \frac{\langle H \rangle}{Vk_BT^2} - \frac{\mu \rho(\infty)}{k_BT^2}.
\end{equation}
\Cref{eq:ctcontact} therefore connects at given $T$ and $\mu$ the contact value of the thermal susceptibility to the mean bulk internal energy per volume, $\avg{H}/V$, and to the bulk density $\rho(\infty)$.

Using a splitting of $\ctr$, we can show that the theorem \eqref{eq:ctcontact} holds separately for the relevant contributions.
To perform the splitting of $\ctr$, we recall its covariance form \eqref{eq:chi-T-cov} and insert the Hamiltonian \eqref{eq:hamiltonian}, which consists of internal, external and kinetic contributions.
For the local thermal susceptibility, this decomposition yields
\begin{align}
  \begin{split}
  \label{eq:ctsplit}
    \ctr &= \frac{1}{k_BT^2}\cov{u(\vec{r}^N),\rhoop{r}}\\
         &\quad + \frac{1}{k_BT^2}\cov{\sum_{i=1}^N V_\mathrm{ext}(\vec{r}_i),\rhoop{r}}\\
         &\quad + \frac{k_BD}{2}\cmr - \frac{\mu}{T}\cmr,
  \end{split}
\end{align}
where the kinetic term has been evaluated with the equipartition theorem,
\begin{equation}
  \begin{split}
    \cov{\sum_{i=1}^N \frac{\vec{p}_i^2}{2m},\rhoop{r}} &= \frac{k_BTD}{2} \cov{N,\rhoop{r}} \\
    &= \frac{(k_BT)^2D}{2} \cmr.
  \end{split}
\end{equation}

When $V_\mathrm{ext}(\vec{r})$ is a hard wall potential, the second term on the right hand side of \cref{eq:ctsplit} vanishes since $V_\mathrm{ext}(\vec{r}) = 0$ for all allowed particle positions.
Then, for the contact location, \cref{eq:ctsplit} can be evaluated by inserting \cref{eq:cmucontact} and \cref{eq:ctcontact}.
A rearrangement yields
\begin{align}
  \label{eq:ctcontactint}
  \chi_{T, \mathrm{int}}(0^+) = \frac{\cov{u(\vec{r}^N),\hat{\rho}(0^+)}}{k_BT^2} = \frac{\langle u(\vec{r}^N) \rangle}{Vk_BT^2},
\end{align}
which constitutes a separate contact theorem for the part of $\chi_T(\vec{r})$ that is generated only by the internal interaction $u(\vec{r}^N)$ of the particles and that we hence indicate as $\chi_{T, \mathrm{int}}(\vec{r})$.

As the thermodynamic relations only hold in the thermodynamic limit, one expects that the above contact theorems are not satisfied in small systems, but converge in simulations for large box sizes, where effects of the wall become negligible for the bulk fluid.
A numerical analysis and validation of the contact theorems is shown in \cref{sec:LJ_Gauss_confinement}.

\subsection{Canonical ensemble}
\label{sec:canonical}
Throughout the previous sections, the grand canonical ensemble has been used to define and investigate fluctuation profiles.
In the following, we show that suitable fluctuation profiles can be derived for the canonical ensemble, where we take as a starting point the Helmholtz free energy $F_N = U_N - T S_N$ as the relevant thermodynamic potential.
All relevant canonical quantities are denoted by a subscript $N$.

We follow a route similar to that taken in \cref{sec:funct-deriv} and consider the functional derivative of $F_N$ (instead of the grand potential $\Omega$ considered in \cref{sec:funct-deriv}) with respect to the external potential $V_\mathrm{ext}(\vec{r})$.
This leads to the fundamental splitting
\begin{equation}
  \label{eq:5}
  \frac{\delta F_N}{\delta V_\mathrm{ext}(\vec{r})} = \frac{\delta U_N}{\delta V_\mathrm{ext}(\vec{r})} - T \frac{\delta S_N}{\delta V_\mathrm{ext}(\vec{r})},
\end{equation}
analogous to \cref{eq:Omega-func}, as the basis for the derivation of canonical fluctuation profiles.

The functional derivatives are evaluated as in \cref{sec:funct-deriv}, but using the canonical trace and probability distribution.
This results in the canonical covariance expressions
\begin{align}
\label{eq:chi-star-canonical}
  \frac{\delta U_N}{\delta V_\mathrm{ext}(\vec{r})} &= \rho_N(\vec{r}) - \beta \covN{H,\rhoop{\vec{r}}} \equiv \chi_{\star, N}(\vec{r}),\\
  \label{eq:chi-T-canonical}
  -T \frac{\delta S_N}{\delta V_\mathrm{ext}(\vec{r})} &= \beta \covN{H,\rhoop{\vec{r}}} \equiv T \chi_{T, N}(\vec{r}),
\end{align}
where $\rho_N(\vec{r})$ is the canonical density profile and $\covN{\cdot, \cdot}$ denotes the covariance \eqref{eq:cov} but build with canonical averages.
As the decomposition \eqref{eq:5} of the free energy does not contain the term $\mu \langle N \rangle$, the local compressibility $\cmr$ cannot be defined as a corresponding functional derivative in the canonical ensemble.

The form of  the canonical covariances \cref{eq:chi-star-canonical,eq:chi-T-canonical} formally coincides with the grand canonical covariance identities \eqref{eq:rhos-cov} and \eqref{eq:chi-T-cov} for $\rhosr$ and $\ctr$, respectively.
However, as different ensemble averages are used in the covariances in \cref{eq:chi-T-canonical,eq:chi-T-func} for $\ctr$ and $\chi_{T, N}(\vec{r})$, respectively, $\ctr$ and $\chi_{T, N}(\vec{r})$ are really distinct from each other.
The same holds analogously for $\rhosr$ and $\chi_{\star, N}(\vec{r})$, and they are therefore also distinct quantities.
This is contrary to observables that are defined as simple averages, such as the density profile $\rhor = \langle \sum_i \delta(\vec{r} - \vec{r}_i) \rangle$, since they generally lose their ensemble-dependence in the thermodynamic limit.
We will investigate the ensemble differences of the fluctuation profiles below in \cref{sec:cgc}.

Similar to \cref{sec:contact_values}, one can proceed with a decomposition of $\chi_{T, N}(\vec{r})$ into
\begin{equation}
  \label{eq:8}
  \begin{split}
    \chi_{T, N}(\vec{r}) &= \frac{1}{k_B T^2} \left[ \covN{u(\vec{r}^N),\hat{\rho}(\vec{r})} \phantom{\sum_{i=1}^N} \right.\\
                         &\qquad \left. + \covN{\sum_{i=1}^N V_\mathrm{ext}(\vec{r}),\hat{\rho}(\vec{r})} \right].
  \end{split}
\end{equation}

In the canonical ensemble, integrals over $\chi_{T, N}(\vec{r})$ vanish due to $N$ being fixed, which can be proven by using the expression in \cref{eq:chi-T-canonical},
\begin{align}
    \intd{r} \chi_{T, N}(\vec{r}) &= \frac{1}{k_B T^2} \intd{r} \covN{H,\hat{\rho}(\vec{r})} \\ &= \frac{1}{k_B T^2} \covN{H,\intd{r} \hat{\rho}(\vec{r})} \\ &= \frac{1}{k_B T^2} \covN{H,N} \\ &= \frac{N}{k_B T^2} \covN{H,1} = 0.
    \label{eq:vanishctc}
\end{align}
This holds separately as well for the internal and external contribution of $\chi_T(\vec{r})$, cf.\ \cref{eq:ctsplit}.

We note that in principle, one could still get the grand canonical $\ctr$ from canonical data by taking the partial derivative route in \cref{eq:chi-T-para}, which states that $\rhor$ must be derived with respect to $T$ along constant $\mu$.
Practically, this is hindered by the fact that the particle number $N$ rather than the chemical potential $\mu$ is controlled in the canonical ensemble, thus requiring some measurement of $\mu$ \cite{Widom1963}.

\section{Simulation results}
\label{sec:simulation}
\subsection{Numerical methods}
As we have shown in \cref{sec:funct-deriv,sec:correlator_expressions} on the basis of the grand canonical ensemble, several possible techniques exist to make the fluctuation profiles accessible in molecular simulations.
We recall that one option is the explicit evaluation of the parametric derivatives of the density profile $\rho(\vec{r})$ via finite difference versions of \cref{eq:chi-mu-para,eq:chi-T-para,eq:rhos}.
For this, one must obtain $\rho(\vec{r})$ for slightly different temperatures $T \pm \Delta T$ and chemical potentials $\mu \pm \Delta \mu$  to yield an estimate for the fluctuation profiles at a target thermodynamic state point $T, \mu$.
This requires carrying out multiple distinct simulations.
Arguably more convenient are the correlator expressions \cref{eq:chi_mu_cov,eq:rhos-cov,eq:chi-T-covS}, which enable the sampling of all fluctuation profiles in a single molecular simulation at fixed temperature $T$ and fixed chemical potential $\mu$.

Recall that the definition \eqref{eq:chi-T-func} of $\chi_T(\vec{r})$ contains thermal quantities due to the appearance of the entropy operator $\hat{S} = -k_B \ln\Psi$.
We reemphasize that nevertheless $\chi_T(\vec{r})$ can be reformulated in equilibrium via \cref{eq:chi-T-cov} as an expression that avoids the explicit evaluation of the partition function or a thermodynamic potential.
The local thermal susceptibility can thus be sampled in a straightforward manner similar to what is possible for $\chi_\mu(\vec{r})$ using only averages that are easily accessible in a standard simulation.

In the following, we present results of grand canonical Monte Carlo simulations \cite{Frenkel2023} both for fluids in bulk and in planar confinement.
We obtain the fluctuation profiles via the correlator sampling scheme.
Apart from the practical superiority in this context, the correlator sampling method also leads to reduced error propagation of compound expressions such as $\chi_T(\vec{r}) / \chi_\mu(\vec{r})$, which will become important below for hard sphere systems.
Nevertheless, we have verified consistency with the partial derivative route.

We examine fluids confined by planar walls perpendicular to the $x$-direction and retain a planar geometry by imposing periodic boundary conditions along the $y$- and $z$-axis.
The walls are modeled by an external potential $V_\mathrm{ext}(x)$ for which two common forms will be considered.
We only give the expressions of $V_\mathrm{ext}(x)$ for the left wall at $x = 0$ in the following.

First, we consider the behavior at a planar hard wall as represented by the external potential
\begin{equation}
  \label{eq:Vext_hard}
  V_\mathrm{ext}(x) =
  \begin{cases}
    0 & x > 0,\\
    \infty & x \leq 0.
  \end{cases}
\end{equation}
Second, a truncated and shifted Lennard-Jones potential is imposed as a representative soft wall given by
\begin{equation}
  \label{eq:Vext_LJ}
  V_\mathrm{ext}(x) = \frac{\varepsilon_w}{4} \left[ \left( \frac{\sigma_x}{x} \right)^9 - 3 \left( \frac{\sigma_x}{x} \right)^3 - \left( \frac{\sigma_x}{x_c} \right)^9 + 3 \left( \frac{\sigma_x}{x_c} \right)^3 \right],
\end{equation}
for $x \leq x_c$ and $V_\mathrm{ext}(x) = 0$ for $x > x_c$ with cutoff distance $x_c$, energy scale $\varepsilon_w$ and interaction length $\sigma_x$.

In the following, the influence of these external potentials on the spatial structure of the fluctuation profiles is examined for systems consisting of hard sphere, Gaussian and Lennard-Jones particles.
We remind the reader that we keep the full temperature dependence of the thermal de Broglie wavelength $\Lambda=\Lambda(T)$.

\subsection{Hard Particles}
\label{sec:hard}

\begin{figure}[htb]
  \centering
  \includegraphics[width=\linewidth]{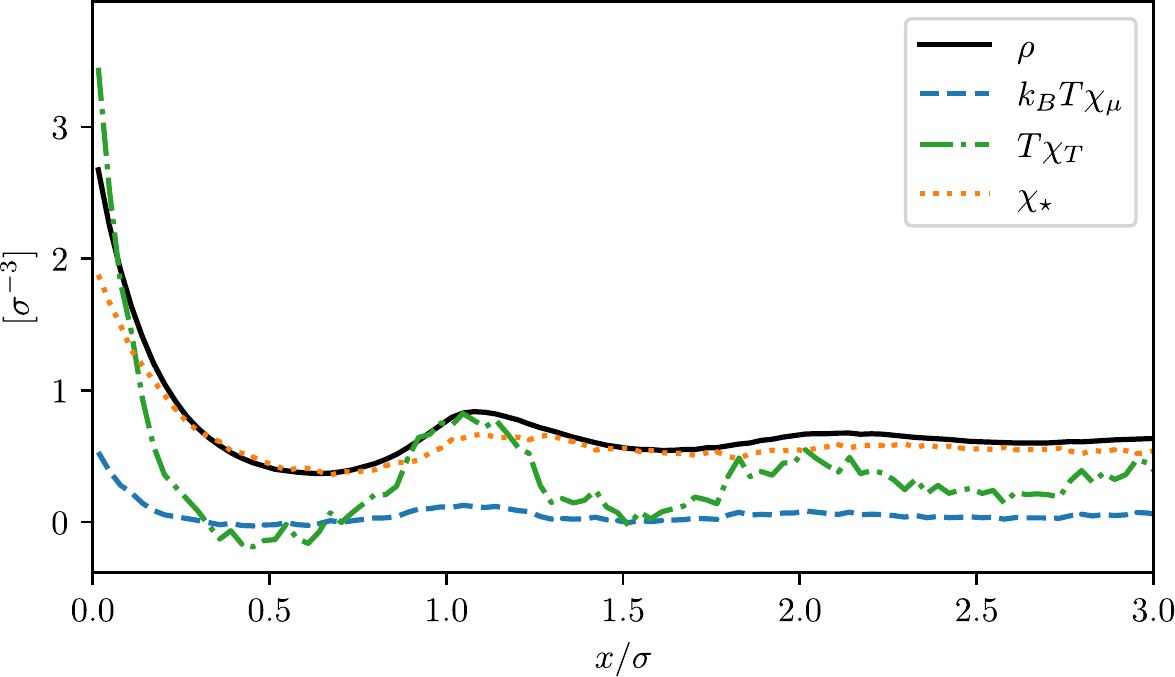}
  \caption{
    The three fluctuation profiles and the density profile for a hard sphere bulk fluid in contact with a hard wall at $\mu = -5.0 k_BT$.
    Layering near the wall is observable in all three fluctuation profiles.
    Local maxima and minima coincide approximately for all fluctuation profiles, although small deviations can still be noticed.
    Interestingly, while $\rho(x) > 0$ is composed of the three profiles, $\chi_T(x)$, $\chi_\mu(x)$ and $\chi_\star(x)$, each one of them does not have to be strictly positive.
    Here, negative values of $\chi_T(x)$ and $\chi_\mu(x)$ occur at $x \approx 0.5 \sigma$.
    Note that $\cmr$ must be positive in bulk due to \cref{eq:chi_mu_bulk_positive}.
  }
  \label{fig:chis_hard}
\end{figure}

\begin{figure}[htb]
  \centering
  \includegraphics[width=\linewidth]{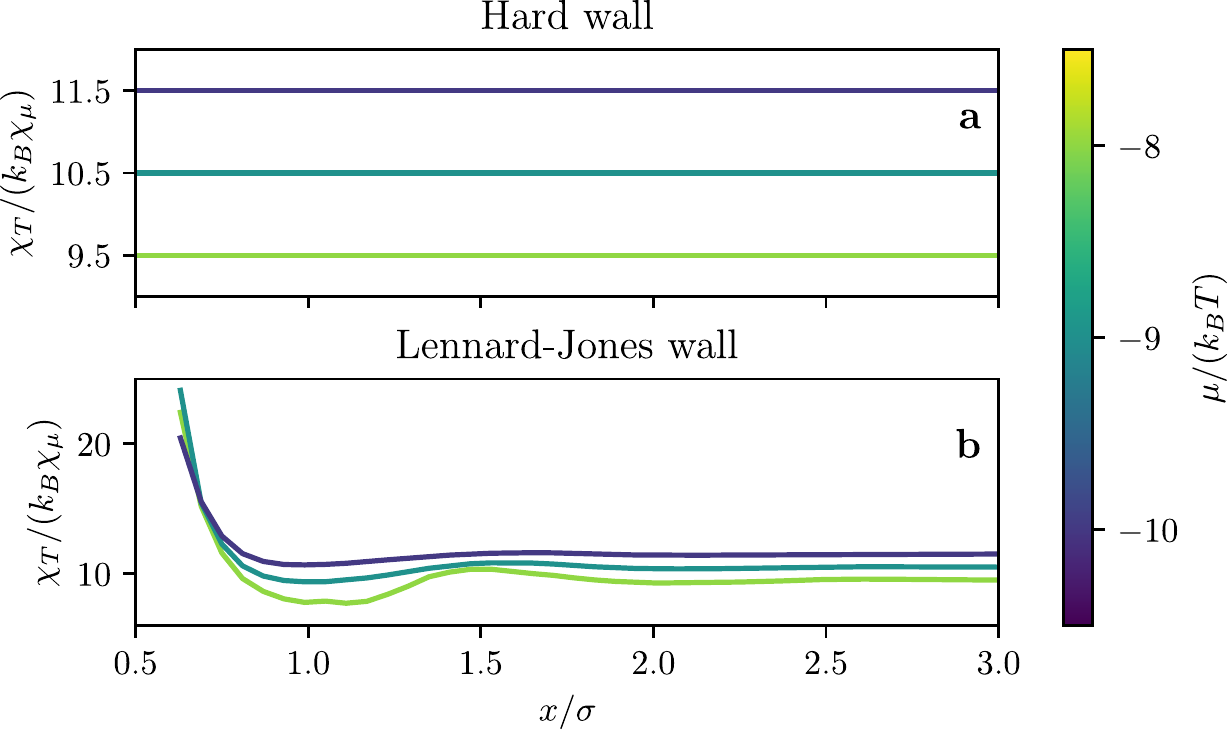}
  \caption{
    The ratio $\chi_T(x) / \chi_\mu(x)$ for the hard sphere fluid in contact with (a) a hard wall and (b) a Lennard-Jones wall is shown for varying chemical potential $\mu / (k_BT) = {-8, -9, -10}$.
    While the fluctuation ratio remains spatially constant in (a) and attains values according to \cref{eq:chi_T_chi_mu_hard}, it shows oscillatory behavior in (b).
    This is due to \cref{eq:chi_T_hard_density_cov}, and shows that non-local effects of a soft external potential can be observed in the fluctuation ratio.
  }
  \label{fig:chi_T_chi_mu_sweep}
\end{figure}

We first turn to the arguably simplest model of a confined fluid, namely hard spheres of diameter $\sigma$ confined by two parallel planar hard walls.
The distance between the walls is chosen sufficiently large such that a mutual influence is negligible and the simulation effectively serves as an approximation of a bulk fluid in contact with a single hard wall.
\Cref{fig:chis_hard} shows the numerical results for all three fluctuation profiles obtained from this simulation.
The quantities $\chi_T(\vec{r})$ and $\chi_\mu(\vec{r})$ show strong oscillatory behaviour near the hard wall.
They even become negative around $x \approx 0.5 \sigma$.

Although the strictly positive density profile $\rhor$ is a linear combination of $\chi_T(\vec{r})$, $\chi_\mu(\vec{r})$ and $\chi_\star(\vec{r})$, the fluctuation profiles do not have to obey the constraint of positivity separately.
For $\cmr$, this stands in contrast to the behaviour of the corresponding bulk value, which is necessarily positive, as shown in section \ref{sec:funct-deriv}.
Negative values can also be seen in results by \citeauthor{Evans2015} \cite{Evans2015} for the local compressibility $\chi_\mu(\vec{r})$ in Lennard-Jones liquids near a solvophilic wall.
Our simulation results show that this effect does not require attractive interactions -- be they from internal or external contributions.
The sign change rather also occurs in an athermal situation where only hard core repulsions exist.

For hard spheres, we have shown in \cref{sec:hard_core} that the fluctuation ratio $\ctr / \cmr$ is a natural quantity to consider, and that this ratio is expected to show markedly different behavior depending on the type of the external potential.
In \cref{fig:chi_T_chi_mu_sweep}, this ratio is shown both for the case of a hard wall as well as for a Lennard-Jones wall given by \cref{eq:Vext_LJ} with $\sigma_x = \sigma$, $x_c = 2 \sigma$ and $\varepsilon_w = k_BT$.
For the hard wall, the ratio $\ctr / \cmr$ remains spatially constant and attains a value consistent with \cref{eq:chi_T_chi_mu_hard}.
Conversely, in the case of the soft Lennard-Jones wall the fluctuation ratio deviates from its constant bulk value and shows oscillatory behavior.

In the vicinity of the wall, the value of the susceptibility ratio increases and resembles the shape of the external potential, reproducing its minimum at $x \approx 1 \sigma$.
Especially for higher packing fractions, i.e.\ for larger values of $\mu$, oscillations that reach a few hard sphere diameters into the bulk fluid can be observed.
This behaviour can be traced back to \cref{eq:chi_T_hard_density_cov}, which shows that effects of the external potential are mediated non-locally via $H_2(\vec{r}, \vec{r}')$ and contribute directly to the fluctuation ratio.

\subsection{Lennard-Jones and Gaussian Particles}
\label{sec:LJ_Gauss_confinement}

\begin{figure}[h]
  \centering
  \includegraphics[width=\linewidth]{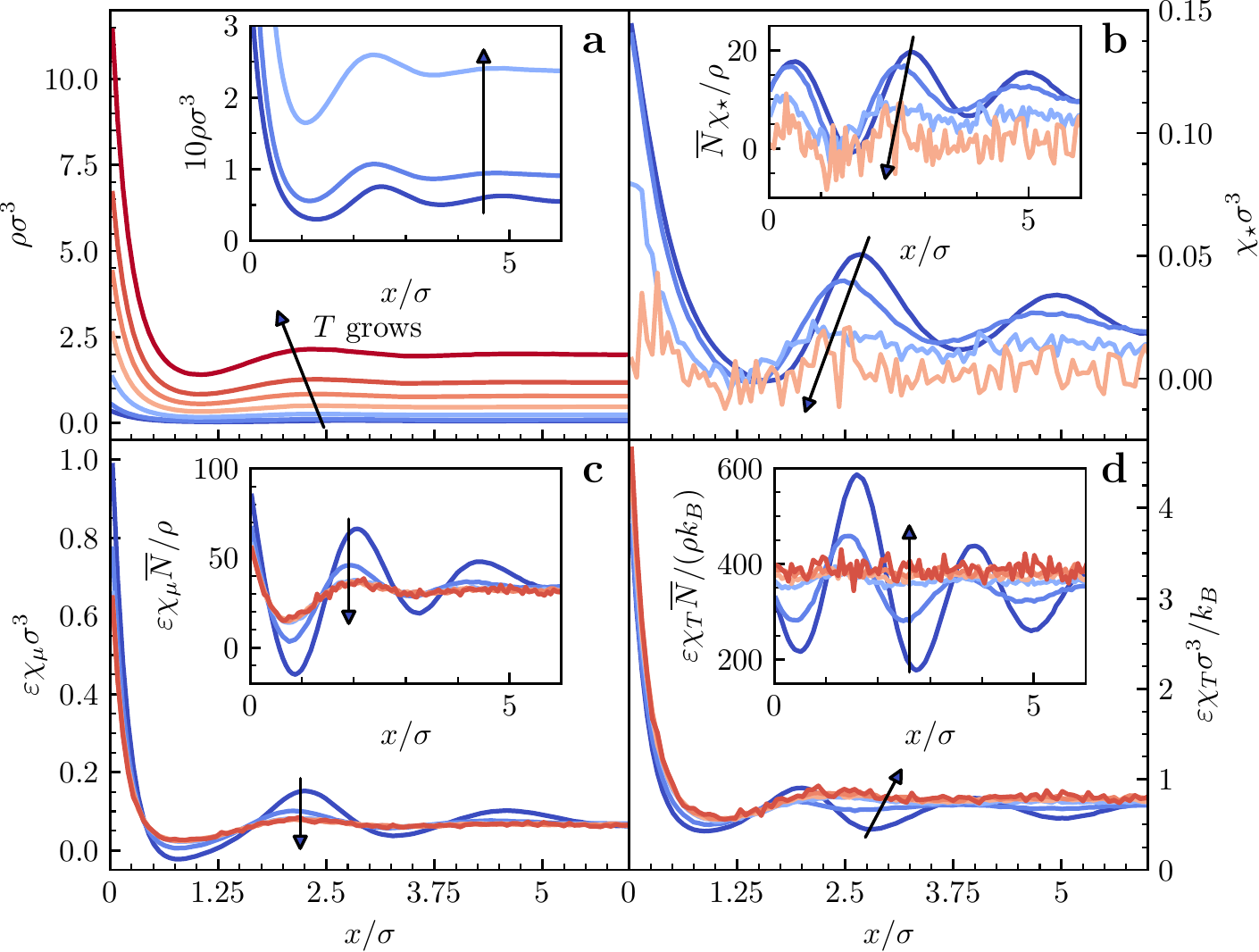}
  \caption{
    Gaussian core model fluid confined between hard walls at $x = 0$ and $12 \sigma$.
    The density profile $\rho(x)$ (a), reduced density $\chi_{\star}(x)$ (b), local compressibility $\chi_{\mu}(x)$ (c), and local thermal susceptibility $\chi_T(x)$ (d) are shown.
    The chemical potential $\mu = 0$ is fixed and the temperature range $k_BT/\varepsilon \in \{ 0.1, 0.3, 0.6, 1.0, 1.5, 2.5 \}$ is considered.
    Arrows and the color gradient from blue to red thereby indicate increasing temperature.
    The inset in (a) shows the density profile for low temperatures in more detail, whereas the insets in (b), (c) and (d) depict the respective fluctuation profiles divided by the normalized density to illustrate structural variations larger than the density.
  }
  \label{fig:gausswallT}
\end{figure}

\begin{figure}[h]
  \centering
  \includegraphics[width=\linewidth]{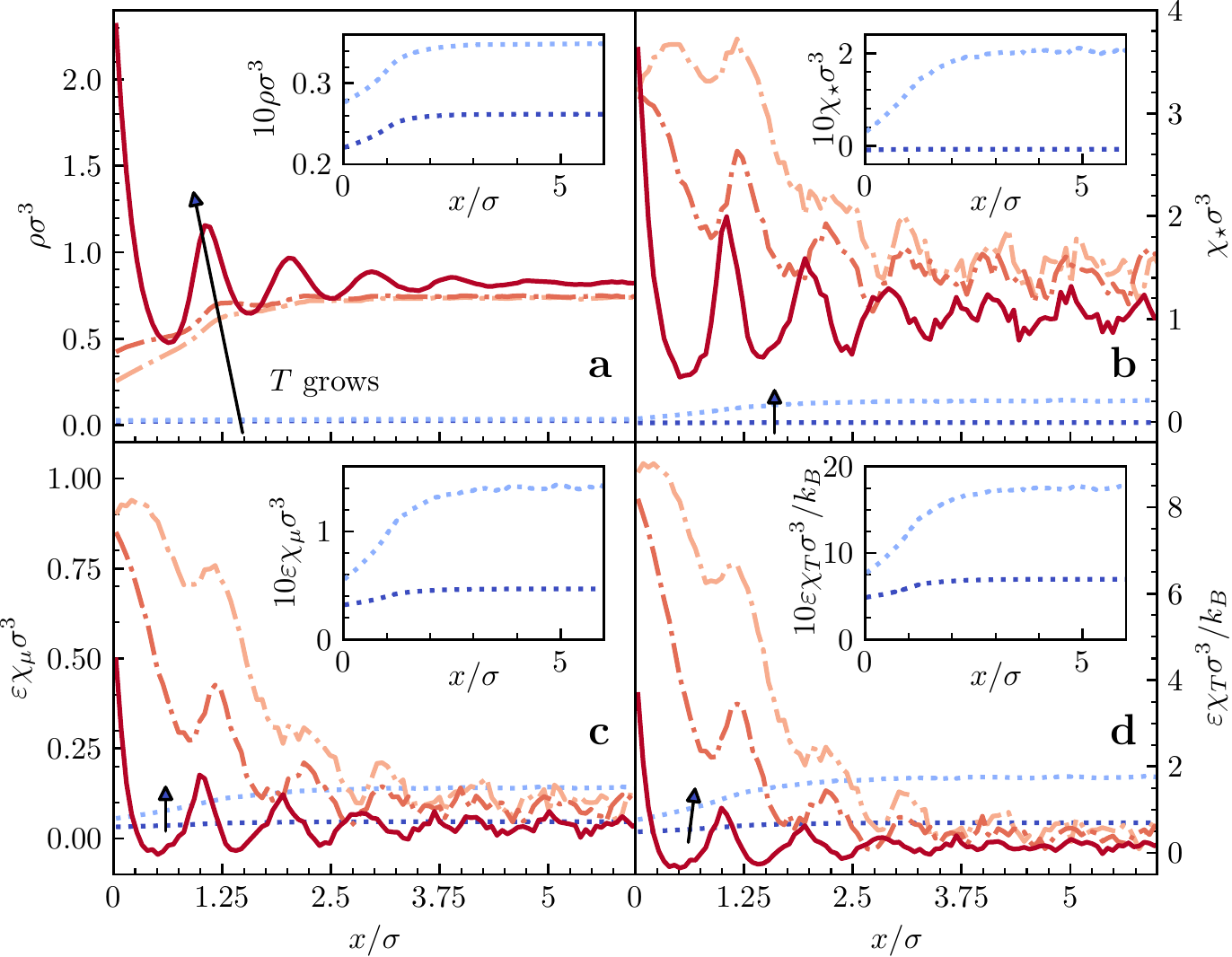}
  \caption{
    Planar system of Lennard-Jones particles confined between hard walls at $x = 0$ and $12 \sigma$.
    The density profile $\rho(x)$ (a), reduced density $\chi_{\star}(x)$ (b), local compressibility $\chi_{\mu}(x)$ (c), and local thermal susceptibility $\chi_T(x)$ (d) are shown for fixed chemical potential $\mu=-11.5\varepsilon$ and varying temperature $k_BT/\varepsilon \in \{0.82, 0.83, 0.84, 0.86, 1.2 \}$.
    Arrows and color gradient from blue to red thereby represent increasing temperature.
    The transition from vapor (dotted) to desorbing liquid (dash-dotted) to adsorbing liquid (solid) can be observed, see Appendix \ref{sec:lambda} for a detailed account of the consequences that arise due to including the full temperature dependence of the thermal wavelength $\Lambda$.
    Insets show the respective fields for the vapor state.
  }
  \label{fig:LJwallT}
\end{figure}

We next turn to the investigation of fluctuation profiles for soft particles, where additional contributions to the Hamiltonian due to a finite interparticle pair potential arise.
As two prototypical microscopic models, we choose i) the Gaussian core model as a purely repulsive particle type that smoothly interpolates from an ideal gas to strongly repulsive interactions and ii) the Lennard-Jones fluid as a standard model fluid with both repulsive and attractive contributions where the presence of the later generates a liquid-vapor phase transition.
In the following, confinement of those fluid models by two parallel planar hard walls is considered.

The local fluctuation profiles are means to probe the vicinity of the thermodynamic state point, cf.\ \cref{sec:funct-deriv}.
Therefore, it is natural to investigate in detail the behavior upon changes in $T, \mu$ in a grand canonical setting.
Depending on the prevailing bulk phase, the potential proximity to bulk and surface phase transitions, and the influence of external potentials, markedly different structure is thereby expected.

For simplicity, we keep the chemical potential $\mu$ fixed and vary the temperature $T$ across a suitable range.
For the Gaussian core model, this enables the analysis of the emergence of structure in the fluctuation fields when gradually decreasing $T$, as the influence of the interparticle potential barrier becomes increasingly dominant.
The energy scale of the Gaussian interparticle interaction potential is $\varepsilon$, the standard deviation is $\sigma$ and the cutoff radius is $3\sigma$.

The results for temperatures ranging from $k_BT / \varepsilon = 0.1$ to $2.5$, and for fixed chemical potential $\mu = 0$ are shown in \cref{fig:gausswallT}.
All three fluctuation profiles are nearly featureless for high temperature.
Such behaviour is expected due to the small influence of the interparticle interactions at high temperature.
For lower temperatures, the gradual formation of oscillations can be seen.
The observed structure of fluctuation profiles for the Gaussian core model in this situation is -- as anticipated -- similar to the structure observed e.g.\ in \cref{fig:chis_hard} for hard spheres.

Next, we consider the Lennard-Jones fluid with energy scale $\varepsilon$, length scale $\sigma$ and cutoff radius $2.5\sigma$, and show results in \cref{fig:LJwallT}.
We proceed similarly to above and investigate a temperature sweep from $k_BT / \varepsilon = 0.82$ to $1.2$.
Within this range, a liquid-vapor bulk phase transition occurs, the effects of which are reflected clearly in the fluctuation profiles.
While the fluctuation profiles in the vapor phase show no significant structure, the sudden appearance of oscillations can be observed when the system transitions to a liquid.
Note that in the density profile, apart from a global jump to larger values due to the increased mean density, very little structural information is revealed as $\rho(\vec{r})$ remains rather smooth.

This discrepancy between an almost featureless density profile and a dominant signal in the fluctuation profiles is especially obvious when surface effects are accounted for.
For temperatures just below the bulk liquid-vapor phase transition, a desorbing fluid is observed, for which a density depletion is visible near the wall.
At even lower temperatures, the liquid becomes adsorbing, i.e.\ the density increases in the vicinity of the walls and shows oscillations.
Especially for the desorbing liquid, the fluctuation profiles attain values that are orders of magnitude larger than in the middle of the box and display oscillating behavior, while the density profile still has no significant structural features.
This phenomenon is known and well studied for the local compressibility \cite{Evans2015}, and in accordance with Ref.\ \cite{Coe2022a,Coe2022,Coe2023}.
We verify here that such an indicative feature is indeed universal to all fluctuation profiles.

A complete set of fluctuation profiles reveals information about general variations of the thermodynamic state point (i.e.\ not just for variations of $\mu$ as in \cite{Evans2015}).
It is therefore natural to consider -- apart from the local compressibility -- the full set of fluctuation profiles to study phase transitions and criticality in confined fluids.
This might therefore be especially important in situations where the considered system is particularly sensitive to changes in a specific thermodynamic variable.
The construction and measurement of the respective fluctuation profile then enables a much more detailed investigation than the density profile alone.
We point the reader again to Ref.\ \cite{Coe2022a} for their investigation of critical drying.

We note that we have identified the bulk behavior by considering the middle of the box.
This assumption is based on the local density approximation (LDA), i.e.\ the premise that locally, for sufficiently small spatial variation of $V_\mathrm{ext}(\vec{r})$, the value of a one-body field attains the value of the corresponding bulk quantity.
That this approximation holds for fluctuation profiles is shown numerically in \cref{sec:lda}.

\begin{figure}[h]
  \centering
  \includegraphics[width=\linewidth]{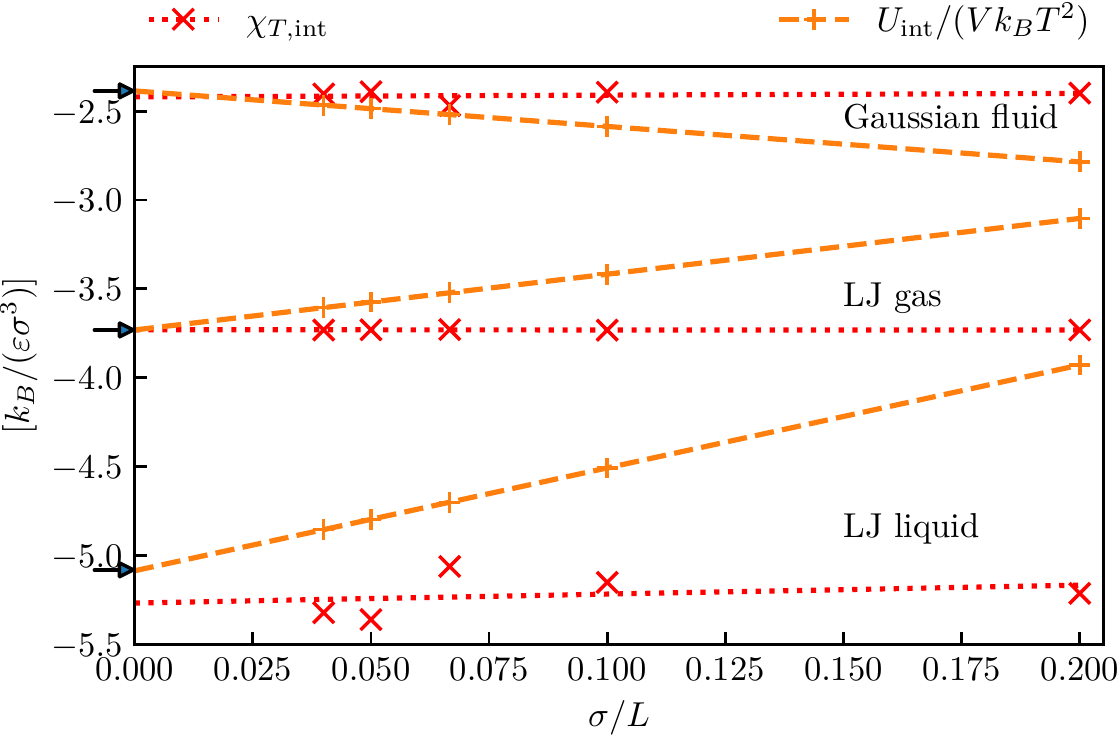}
  \caption{
    Contact value $\chi_{T,\mathrm{int}}(0^+)$ (red crosses) and mean internal interaction energy per volume (orange pluses) in a Lennard-Jones liquid (bottom), Lennard-Jones gas (middle) and Gaussian fluid (top) as a function of system size.
    Linear fits to the data are included as a guide to the eye.
    Results for the Lennard-Jones gas are scaled up by a factor of 2000.
    Arrows on the left side indicate the mean internal energy per volume in a completely periodic bulk system with $L=25\sigma$.
  }
  \label{fig:contact}
\end{figure}

In \cref{sec:contact_values}, we have proven that contact theorems cannot only be derived for $\cmr$ as already done in \cite{Evans2015}, but also for the local thermal susceptibility $\ctr$.
As a result, the contact values of the constituent parts of $\ctr$ are intricately linked to contributions to the global internal energy of the system.

In the following, we demonstrate the validity of the contact theorem \eqref{eq:ctcontact} numerically for the Lennard-Jones vapor and liquid as well as for the Gaussian core model fluid.
To eliminate finite size effects, we perform simulations for a range of box sizes with fixed cross-section $6\sigma \times 6\sigma$ but variable length $L$ and extrapolate $\sigma/L \rightarrow 0$ by fitting a linear function to the contact values thus obtained.
Additionally, bulk simulations are conducted for comparison to the respective bulk value that the theorem connects to the contact values.
Due to the splitting of $\ctr$ according to \cref{eq:ctsplit} and the validity of the contact theorem \eqref{eq:6} for $\cmr$, we focus here on the contact theorem \eqref{eq:ctcontactint} for the only nontrivial contribution $\chi_{T, \mathrm{int}}(0^+)$.
The results are shown in \cref{fig:contact}, and it is evident that the contact theorem \eqref{eq:ctcontactint} indeed holds in all considered cases.
Within numerical accuracy the contact values $\chi_{T, \mathrm{int}}(0^+)$ do not change with the size of the simulation box, which means that even for very small boxes one can estimate the total internal energy of a bulk fluid very accurately.

\subsection{Comparing Canonical and Grand Canonical Systems}
\label{sec:cgc}

\begin{figure}[h]
  \centering
  \includegraphics[width=\linewidth]{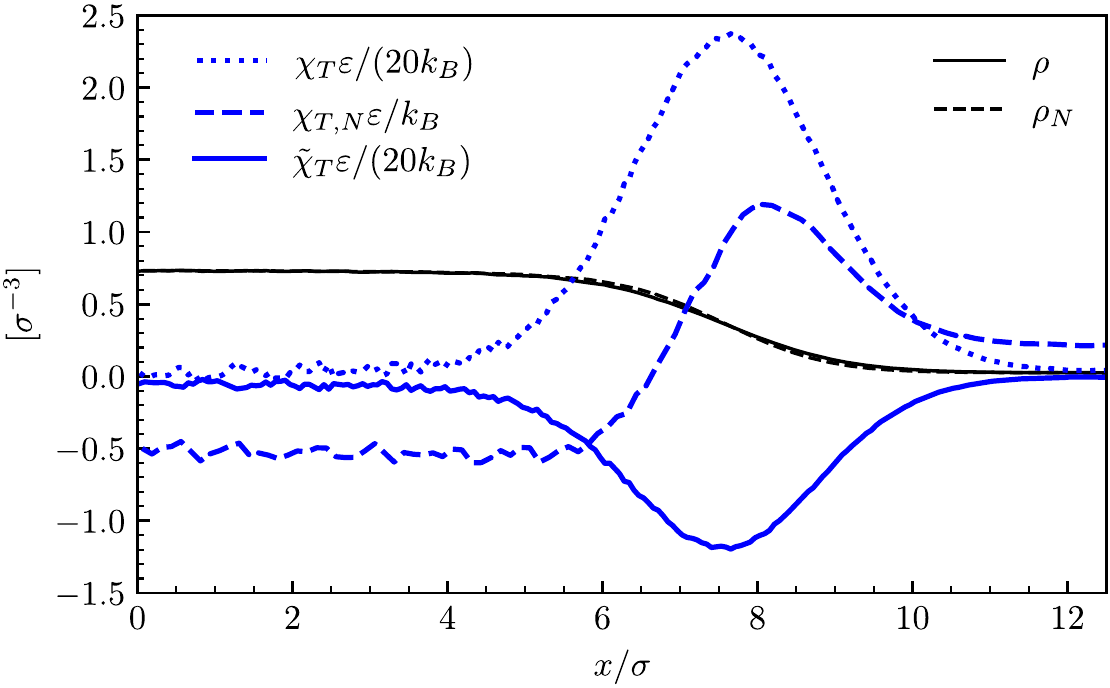}
  \caption{
    Comparison of $\rho(x)$ (black lines) and the thermal susceptibility (blue lines) in canonical (dashed lines) and grand canonical (solid lines) systems of Lennard-Jones particles with liquid-vapor coexistence stabilized by a sinusoidal external potential with a small amplitude of $0.2\varepsilon$.
    We show the full canonical and grand canonical profiles $\chi_{T, N}(x)$ and $\chi_T(x)$ as well as the contribution $\tilde{\chi}_T(x)$ in the grand canonical system that arises solely due to potential energy contributions, see \cref{eq:ctsplit}.
    Both systems are simulated in a box of size $25 \times 6 \times 6 \sigma^3$ at $k_BT = 0.85\varepsilon$.
    The chemical potential in the grand canonical system is set to $\mu=-11.8\varepsilon$, which we have taken as an empirical estimate for the coexistence value at the specified temperature and which leads to a stable liquid-vapor interface in the considered case.
    In the canonical system, $N = 401$ is chosen to match the mean density of the grand canonical system.
  }
  \label{fig:CGCcoex}
\end{figure}

As laid out in \cref{sec:canonical}, fluctuation profiles can be defined canonically, analogous to the procedure based on the grand canonical ensemble, by considering a splitting of the appropriate thermodynamic potential.
We demonstrated that, when taking the covariance route, this yields distinct quantities that differ from the grand canonical counterpart.
Still, the canonical profiles may share the same qualitative behaviour in characterizing density variations as a response to changing the thermodynamic state point and thus, e.g., indicate proximity to phase transitions and critical phenomena.

Recall that it is still possible to obtain ensemble-invariant fluctuation profiles by resorting to the partial derivative route, cf. \cref{eq:chi-mu-para,eq:chi-T-para}, where one must then pay attention to the variables that shall be kept constant (which is cumbersome, but technically possible, in ensembles where those quantities are dependent variables).
For the canonical ensemble, apart from the reduced density $\rhosc$ the thermal fluctuation profile $\ctc$ emerges from the splitting of the Helmholtz free energy $F_N = U_N - TS_N$.
Here we focus on the Lennard-Jones system near liquid-vapor coexistence.
The spatial coexistence is stabilized by a weak sinusoidal external potential in both the canonical and the grand canonical case, separating the system into a liquid and a vapor region.

In order to be able to compare fluctuation profiles of a canonical and grand canonical system, we use the same interaction potential, impose the same temperature and match the mean density of the grand canonical system to the canonical density.
In the case of small particle numbers where ensemble differences are significant, it is generally not possible to obtain the same density profile in both systems if the same external potential is employed.
However, in the thermodynamic limit, ensemble differences in $\rhor$ vanish, such that in the finite simulated systems we are able to achieve nearly the same density profile in sufficiently large canonical and grand canonical simulation domains.
For approaches that relate canonical and grand canonical DFT to each other, see Refs.\ \cite{Gonzalez1997,White2002,Gonzales2000,MS2014}.

We note that $\ctr$ in the grand canonical ensemble consists of multiple contributions as shown in \cref{eq:ctsplit}.
A similar splitting can be performed for the canonical $\ctc$.
To facilitate a concise comparison, we show both the full thermal susceptibility profiles $\ctc$ and $\ctr$ in the canonical and grand canonical ensemble as well as only the constituent of $\ctr$ that arises from contributions of the potential energy, which we denote by $\tilde{\chi}_T(\vec{r})$ in the following.
Note that formally, the correlator expressions for $\ctc$ and $\tilde{\chi}_T(\vec{r})$ contain the same terms and only differ in the form of the ensemble average, cf.\ \cref{eq:ctsplit,eq:8}.
\Cref{fig:CGCcoex} shows these profiles as well as the density in the phase-separated Lennard-Jones systems.

As expected, the densities of the canonical and grand canonical systems almost coincide.
The local thermal susceptibilities $\ctc$ and $\ctr$ on the other hand are very different, which manifests the anticipated ensemble difference.
Nevertheless, for both systems, we expect the extrema of the local thermal susceptibility to occur in regions of substantial particle fluctuation.
This can be verified in \cref{fig:CGCcoex}, and showcases that observables defined by the formalism of \cref{sec:theory} are indeed capable of identifying such regions.
We observe a larger magnitude of the grand canonical thermal susceptibility $\ctr$ as compared to the canonical $\ctc$, which can be attributed to the variability of the particle number $N$ in the grand ensemble.
Note that this also leads to an indirect effect which results in larger fluctuations of the internal interaction potential $u(\vec{r}^N)$.
To further interpret the results in the grand canonical case, it is instructive to consider, say, a growth of the liquid region, thereby shifting the location of the interface and leading to an increase of the total particle number of the system.
As a result, more particles within the fluid phase reside in the well of the pair potentials of their neighbors and decrease the internal energy $u(\vec{r}^N)$.
Therefore, fluctuations in the particle number are anti-correlated with the value of the total internal energy and a significant negative peak forms in $\tilde{\chi}_T(\vec{r})$ in the vicinity of the interface.
In the total profile $\ctr$, this effect is counteracted by the additional terms proportional to $\cmr$, see \cref{eq:ctsplit}.
In the canonical case, fluctuations in the particle number cannot occur, thus leading to a reduced variation of $u(\vec{r}^N)$.
Nevertheless, the dominant scenario for particles within the interface is the escape to the gas phase, thereby increasing the potential energy of the system.
Hence, one observes a positive peak of $\ctc$ at the interface, which is slightly shifted towards the gas phase.
In total, while $\ctr$ and $\ctc$ differ quantitatively, they are equally suitable for localizing fluctuations that are consistent with the considered ensemble.
This information cannot be easily determined by analyzing the density profile alone.

\section{Conclusion and outlook}
\label{sec:conclusion}
We have described different statistical mechanical approaches to the derivation of a complete set of fluctuation profiles.
For the grand canonical ensemble, this set comprises the local compressibility $\cmr$, the local thermal susceptibility $\ctr$ and the reduced density $\rhosr$.

The first way of defining these fields is based on the Legendre structure of the thermodynamic potential as laid out in \cref{sec:funct-deriv}.
A seperate evaluation of functional derivatives that emerge from the splitting \cref{eq:Omega-func} of $\Omega$ reveals generator expressions for fluctuation profiles.
We get those definitions in the grand canonical ensemble by utilizing the fundamental relation $\Omega = U - TS - \mu \langle N \rangle$ of the grand potential, which yields the generator \cref{eq:chi-mu-func,eq:chi-T-func,eq:rhos-generator} for $\cmr$, $\ctr$ and $\rhosr$, respectively.
This constitutes a formalism for the derivation of suitable fluctuation fields also for the canonical ensemble in \cref{sec:canonical}, and it shows that those fields -- similar to the density profile $\rho(\vec{r})$ -- attain a fundamental status due to their connection to thermodynamics.

By an exchange of the order of functional and thermodynamic parametric derivatives, we have constructed an equivalent definition of fluctuation profiles via the parametric derivatives \eqref{eq:chi-mu-para} and \eqref{eq:chi-T-para} of the density profile.
This reveals an intuitive interpretation of the fluctuation profiles as indicators of local susceptibilities of the density upon changes in $\mu, T$, thereby probing the vicinity of the thermodynamic state point.
This route also provides a straightforward way of accessing these fields from simulation data: by the evaluation of finite differences of the density profile, obtained from multiple simulations close to the target thermodynamic state point, one is able to obtain numerical results for $\cmr$, $\ctr$ and $\rhosr$.

A simpler numerical alternative for the evaluation of fluctuation profiles opens up when considering the third possible definition based on covariances as in \cref{sec:correlator_expressions}.
Explicit evaluation of the functional generators is possible (as the underlying distribution function is known in equilibrium), and yields \cref{eq:chi_mu_cov,eq:rhos-cov,eq:chi-T-cov} as covariance expressions for $\cmr$, $\rhosr$ and $\ctr$, respectively.
As all fields obtained from such functional generator expression can be written as covariances, their role as spatially localized measures of fluctuations is again manifest.
Also, the covariance route provides a convenient way of numerically determining fluctuation profiles, as only a single simulation run at fixed thermodynamic state point is needed.

Having multiple definitions of the fluctuation profiles at hand, one can analytically evaluate them for the ideal gas, as done in \cref{sec:ideal-gas}.
From there, for arbitrarily interacting systems, inhomogeneous Ornstein-Zernike equations were derived in \cref{sec:OZ} for the excess (i.e.\ over ideal) parts of $\cmr$ and $\ctr$, which are of simpler one-body type as compared to the standard inhomogeneous OZ equation.
If appropriate closure relations could be found in future work, one might get a feasible scheme for obtaining the inhomogeneous fluctuation profiles -- akin to liquid integral equation theory \cite{Hansen2013} -- from these Ornstein-Zernike equations.
A more immediate application arises by considering the fluctuation OZ equations as easily accessible sum rules, which could prove to be useful in the development and verification of novel numerical methods and machine learning techniques \cite{Heras2023}.

We specialized to the hard sphere model fluid in \cref{sec:hard_core} and derived contact theorems for the fluctuation profiles of fluids in contact with a hard wall in \cref{sec:contact_values}.
The derivation of a contact theorem for the local compressibility $\cmr$ has been done in \cite{Evans2015}; we complement this derivation with one for the local chemical susceptibility $\ctr$.
Thereby, a natural splitting of $\ctr$ into additive contributions inherited from the Hamiltonian is found, and we show that hard wall contact theorems can be derived.

Next we turned to numerical investigation via grand canonical Monte Carlo simulations.
Our focus was on showcasing fluctuation profiles for a range of different fluids in confinement, and to verify some of the predicted properties such as the sensitivity to phase transitions and the contact theorems.
To prove the universal accessibility, we have considered hard core, Gaussian core and Lennard-Jones model fluids as three fundamentally different particle types.
We discriminated between the behavior found at hard walls and at soft walls of Lennard-Jones type.

Especially for fluids undergoing phase transitions or showing surface effects, we saw that fluctuation profiles provide additional structural information that can hardly be revealed on the basis of the density profile alone.
The fluctuation profiles are therefore suitable indicators of such phenomena.
While this has also been described in other works for the local compressibility $\cmr$, we emphasize that a complete set of fluctuation profiles gives additional information of the vicinity of the thermodynamic state point, and is therefore natural to consider in its entirety.
This might be especially important if the system under consideration is predominantly affected by changes in one specific thermodynamic variable.

Lastly, a numerical comparison of canonical and grand canonical systems was given for a spatially phase-separated Lennard-Jones fluid in \cref{sec:cgc}.
While the relevant fluctuation profiles do not coincide quantitatively as they carry an intrinsic ensemble difference (cf.\ \cref{sec:canonical}), their qualitative behavior matches and can be deemed universal in that it localizes regions of large fluctuations (in the presented case the liquid-vapor interface).

As an outlook on possible future work, we mention the investigation of fluctuation profiles for more sophisticated particle models and especially for water due to its crucial role in nature.
For the latter, a variety of atomistic \cite{KadaoluwaPathirannahalage2021} and coarse-grained \cite{Lu2014} interaction potentials exist, and one could assess the resulting fluctuation profiles as an additional criterion of their accuracy.
We note that density fluctuations near a curved hydrophobic solute have recently been investigated for a monatomic water model \cite{Molinero2009} via the local compressibility by \citeauthor{Coe2022a} \cite{Coe2022a}.
The generalization to nonequilibrium interfacial thermodynamics (see e.g.\ Refs.\ \cite{Stierle2021,Rauscher2022} for recent work) is an open question for future work as is the investigation of thermal Noether invariance~\cite{Hermann2022,Hermann2022a,Hermann2022b,Sammueller2023a} on local fluctuations.

Additionally, in this work, we focused on the liquid gas regime and refrained from considering the freezing transition, which occurs e.g.\ for the Gaussian core model both into a bcc as well as an fcc crystal \cite{Prestipino2005}.
Still, the fluctuation profiles could show intricate behavior in this case as well, and may be considered as precursors of crystallization in the future.

\section*{Data availability statement}
The data cannot be made publicly available upon publication because they are not available in a format that is sufficiently accessible or reusable by other researchers. The data that support the findings of this study are available upon reasonable request from the authors.

\begin{acknowledgments}
  We thank Bob Evans, Mary Coe, Peter Cats, and Daniel de las Heras for useful discussions.
  This work is supported by the German Research Foundation (DFG) via project number 436306241.
\end{acknowledgments}

\section*{Conflict of interest}
There are no conflicts to declare.

\bibliography{paper.bib}

\appendix

\section{Functional derivatives}
\label{sec:func-derivatives}
In \cref{sec:correlator_expressions}, covariance forms of the fluctuation profiles $\cmr$, $\ctr$ and $\rhosr$ are derived by explicit evaluation of the functional derivatives of $\avg{N}$, $S$ and $U$ \wrt $\Vr$ in \cref{eq:Omega-func}.
In this section, we evaluate recurring terms that emerge within this derivation.
First, we consider the functional derivative of the Hamiltonian $H$ as given in \cref{eq:hamiltonian} with respect to $\Vr$, which yields the density operator,
\begin{equation}
  \label{eq:func-H}
  \funcVr{H} =  \sum_{i=1}^N \delta(\vec{r}_i-\vec{r}) = \rhoop{r}.
\end{equation}
Next, we calculate the variation of the partition sum $\Xi$ \wrt $\Vr$ where we use \cref{eq:func-H} to obtain
\begin{equation}
  \label{eq:func-Xi}
  \begin{split}
    \funcVr{\Xi} &= -\beta \Tr \exp{-\beta \left( H - \mu N \right)} \rhoop{r} \\
    &= -\beta \rhor \Xi,
  \end{split}
\end{equation}
which gives a term proportional to the density profile $\rhor = \Tr \rhoop{r} \Psi$.
Note that \cref{eq:func-Xi} can also be obtained by using \cref{eq:Omega-rho}.
Finally, \cref{eq:func-Xi} is used to differentiate the Boltzmann distribution function $\Psi$, \cref{eq:prob-distribution}, which yields
\begin{equation}
  \label{eq:func-psi}
  \funcVr{\Psi} = -\beta \left[ \rhoop{r} - \rhor \right] \Psi.
\end{equation}

\section{Temperature-dependence of thermal wavelength}
\label{sec:lambda}

\begin{figure}[htb]
  \centering
  \includegraphics[width=\linewidth]{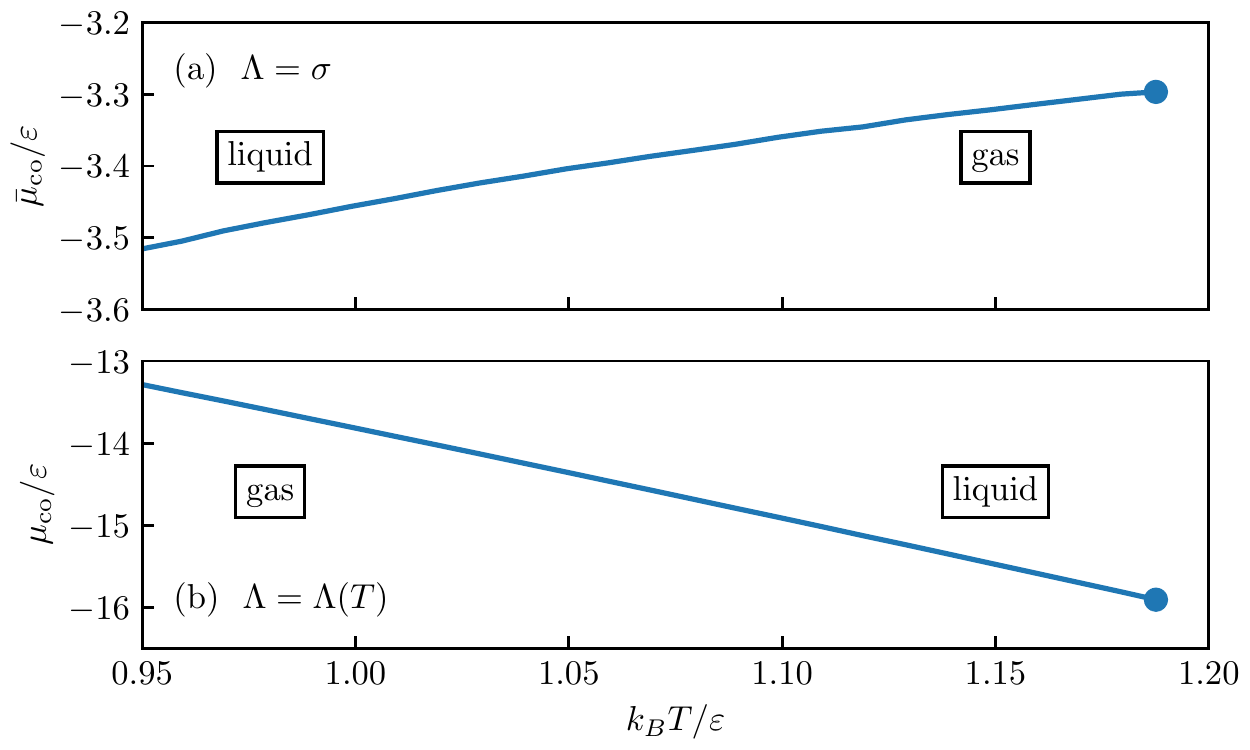}
  \caption{
    We show how the $\mu$-$T$ coexistence curve of a truncated and unshifted Lennard-Jones fluid (taken from Ref.~\onlinecite{Wilding1995}) transforms when including the temperature-dependence of the thermal wavelength $\Lambda(T)$.
    In panel (a), the original data for the coexistence curve $\bar{\mu}_\mathrm{co}(T)$ is reproduced up to the critical point (blue circle) for the usual convention of setting $\Lambda = \sigma$.
    The transformation \eqref{eq:mu_transform} is applied in panel (b), yielding the coexistence curve $\mu_\mathrm{co}(T)$ where the full temperature dependence $\Lambda(T)$ is now taken into account.
    This leads to an inversion of behavior when increasing the temperature $T$ at fixed chemical potential (in the respective sense).
  }
  \label{fig:coexistence_convention}
\end{figure}

The explicit consideration of the temperature dependence of the thermal wavelength $\Lambda$ leads to a transformation of the chemical potential as compared to the usual convention of setting $\Lambda = \sigma$.
This is due to a transformation of the activity $\exp{\beta \mu N}$ when evaluating the kinetic part of the grand canonical average \eqref{eq:average}.
Insertion of \cref{eq:prob-distribution,eq:trace} yields
\begin{equation}
  \avg{A} = \sum_{N=0}^\infty \frac{\mathrm{e}^{\beta \mu N}}{N! \Lambda^{DN}} \intdN{r} \frac{\mathrm{e}^{-\beta H}}{\Xi} A,
\end{equation}
for phase space functions $A\left(\vec{r}^N\right)$ that do not depend on the momenta $\vec{p}^N$; we recall the form $\Lambda = h / \sqrt{2 \pi m k_\mathrm{B} T}$ of the thermal wavelength.

To identify the transformation law, we focus on the prefactor of the spatial integrals and observe
\begin{equation}
  \frac{\mathrm{e}^{\beta \mu N}}{\Lambda^{DN}} = \exp{\beta N \left[\mu - k_B T D \ln\left(\frac{\Lambda}{\sigma}\right)\right]} \sigma^{-DN}.
\end{equation}
Therefore, when setting $\Lambda = \sigma$, the logarithm vanishes and one recovers the usual convention for measurements of the chemical potential.
On the other hand, taking the full temperature dependence of $\Lambda(T)$ into account leads to an additive term in the chemical potential according to
\begin{equation}
  \label{eq:mu_transform}
  \mu = \bar{\mu} - k_B T D \ln\left(\frac{\Lambda(T)}{\sigma}\right),
\end{equation}
where we now identify $\bar{\mu}$ with the chemical potential in the case $\Lambda = \sigma$.

This mere change in the convention of defining a chemical potential leaves the physics unchanged but has ramifications for the interpretation of results.
In particular, when considering the Lennard-Jones fluid in \cref{sec:simulation}, the liquid-vapor coexistence curve assumes a different trend.
We illustrate this in \cref{fig:coexistence_convention}, where we show the coexistence curve of the truncated ($r_c = 2.5 \sigma$) and unshifted Lennard-Jones fluid, which we reproduced from Ref.~\onlinecite{Wilding1995}.
The application of the above transformation results in an inversion of the phase transition when changing the temperature at constant chemical potential in the respective convention.
Although we consider the shifted Lennard-Jones fluid in the main text, we expect this qualitative trend to persist, hence clarifying the observed behavior in \cref{fig:LJwallT}.

\section{Ideal gas}
\label{sec:ideal-gas}
In a grand canonical ideal gas, the density profile is known analytically,
\begin{equation}
  \label{eq:density-ideal-gas}
  \rhor = \Lambda^{-D} \exp{ - \beta (\Vr - \mu) }.
\end{equation}
Therefore, we can evaluate the derivatives \wrt $T$ and $\mu$ explicitly to obtain analytic expressions for the ideal fluctuation profiles.

We start with the differentiation of $\rhor$ \wrt $\mu$ using \cref{eq:density-ideal-gas} and obtain straightforwardly:
\begin{equation}
  \label{eq:derivative-density-mu}
  \cmidr = \left.\frac{\partial \rhor}{\partial \mu}\right|_{T} = \beta \rhor.
\end{equation}
Hence, the ideal local compressibility is simply proportional to the density profile.

To obtain an ideal expression $\ctidr$, we proceed similarly by differentiating \cref{eq:density-ideal-gas} with respect to $T$.
Within the following derivation, the temperature dependence of the thermal wavelength $\Lambda = h\sqrt{2\pi m k_\mathrm{B}T}$ is considered explicitly via \cref{eq:thermal_wavelength_derivative}.
This yields
\begin{align}
  \label{eq:derivatives of density of ideal gas T 1}
  \ctidr &= \left.\frac{\partial \rhor}{\partial T}\right|_{\mu} \\
  \label{eq:derivatives of density of ideal gas T 2}
       &= \frac{\beta}{T} \left(\frac{D}{2} k_B T + (\Vr - \mu)\right) \rhor \\
  \label{eq:derivatives-density-T}
       &= \frac{1}{T} \left( \frac{D}{2} - \ln(\Lambda^D \rho(\vec{r})) \right) \rho(\vec{r}),
\end{align}
where we have used \cref{eq:density-ideal-gas} in the last step to express $\ctr$ solely in terms of the density profile.

Using \cref{eq:rhos} and the ideal expressions \eqref{eq:derivative-density-mu} and \eqref{eq:derivatives-density-T}, we further derive the ideal reduced density
\begin{align}
  \rhosidr &= \rhor - \mu \cmidr - T \ctidr \\
  \label{eq:rhos-final}
  \begin{split}
    &= - \cmidr \Vr + \left(1 - \frac{D}{2}\right) \Lambda^{-D} \\
    &\quad \times\exp{ {- \frac{\ctidr}{k_B \cmidr} + \frac{D}{2}}}.
  \end{split}
\end{align}
To provide additional insight into \cref{eq:rhos-final}, we consider its exponent
\begin{equation}
\label{eq:cm-ct-quotient}
  - \frac{\ctidr}{k_B \cmidr} + \frac{D}{2} = - \beta (\Vr - \mu),
\end{equation}
where we used \cref{eq:derivative-density-mu,eq:derivatives of density of ideal gas T 2}.
By comparison with the ideal density \cref{eq:density-ideal-gas}, we can write
\begin{equation}
\label{eq:rho-chis}
   \rhor = \Lambda^{-D} \exp{ {- \frac{\ctidr}{k_B \cmidr} + \frac{D}{2}} },
\end{equation}
and express the ideal density profile $\rhor$ solely in terms of $\cmidr$ and $\ctidr$.
This enables us to reformulate \cref{eq:rhos-final} as
\begin{align}
\label{eq:rhos-ideal-rho}
  \rhosidr &= \rhor - \left(\beta \Vr + \frac{D}{2}\right) \rhor.
\end{align}

\section{Local Density Approximation}
\label{sec:lda}

\begin{figure}[htb]
  \centering
  \includegraphics[width=\linewidth]{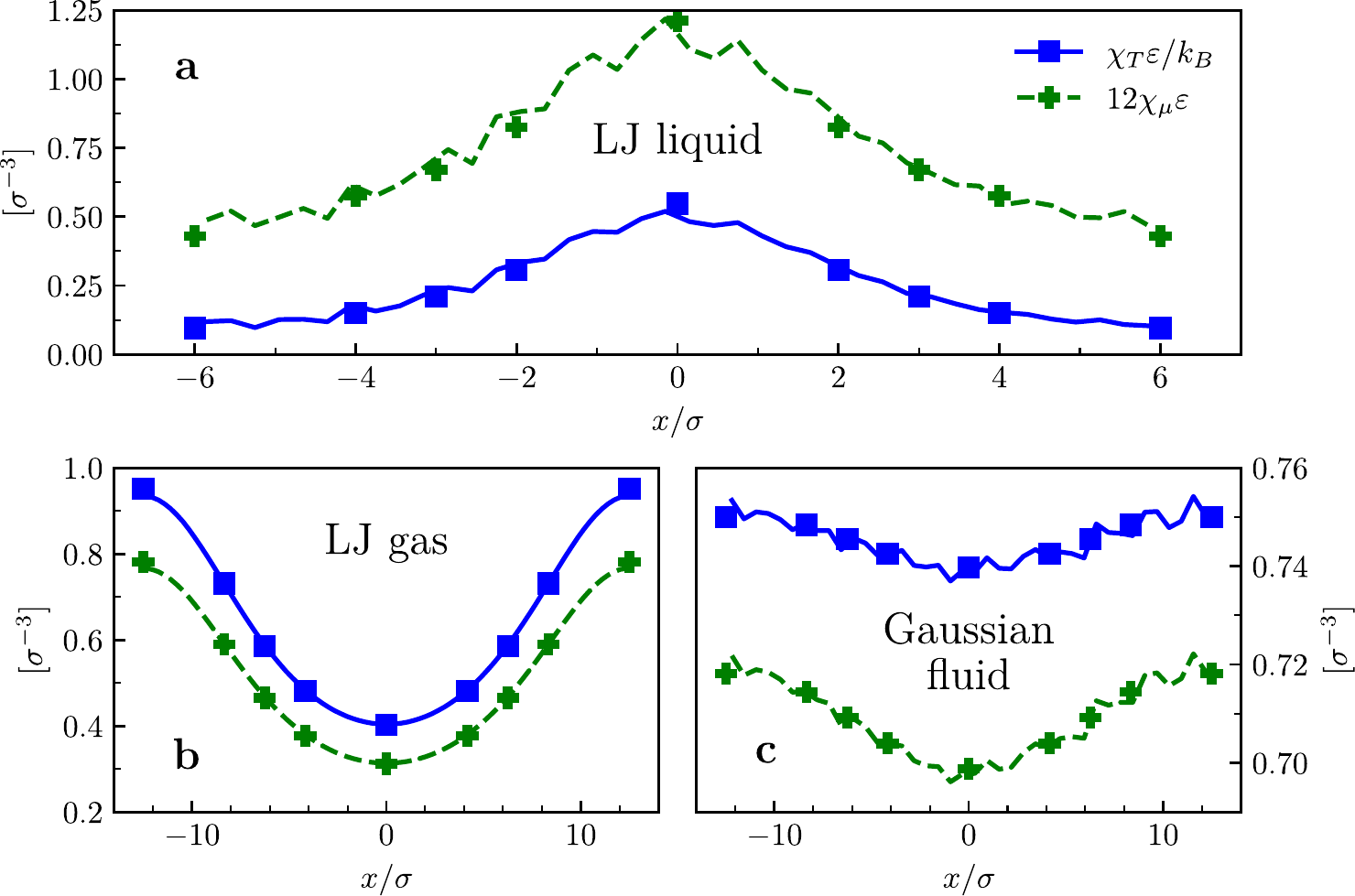}
  \caption{
    Local density approximation for $\chi_T(x)$ (blue squares, solid line), and $\chi_\mu(x)$ (green pluses, dashed line) in a bulk system with external potential \cref{eq:VextLDA}.
    Lines depict inhomogeneous simulation data while symbols are values obtained via LDA.
    We show (a) a Lennard-Jones liquid in a system of size $12 \times 6 \times 6 \sigma^3$ at $\mu=-11\varepsilon$ and $k_B T=0.9\varepsilon$ with $\alpha=1\varepsilon$, (b) a Lennard-Jones gas in a system of size $25 \times 6 \times 6 \sigma^3$ at $\mu=-12\varepsilon$ and $k_B T=0.85\varepsilon$ with $\alpha=0.2\varepsilon$, and (c) a Gaussian fluid in a system of size $25 \times 6 \times 6 \sigma^3$ at $\mu=-4\varepsilon$ and $k_B T=0.5\varepsilon$ with $\alpha=0.2\varepsilon$.
  }
  \label{fig:LDA}
\end{figure}

The local density approximation (LDA) is commonly used in the theory of inhomogeneous liquids \cite{Hansen2013,Elkamel1991,Marconi1991}.
One can qualitatively suppose that LDA is also a viable approximation for fluctuation profiles by comparison of the behavior of bulk and confined systems.
This is not clear a priori, as the covariance expressions contain thermodynamic quantities (e.g.\ $S$, $N$, $U$).

To carry out a quantitative comparison, we impose a modulated external potential
\begin{equation}
\label{eq:VextLDA}
\V(x) = \alpha\cos \left( \frac{2\pi x}{L} \right),
\end{equation}
in a bulk system of size $L \times 6\sigma \times 6\sigma$ where $\alpha$ scales the amplitude of the modulation.
We then compare the obtained profiles at position $x$ with bulk systems that are simulated at the same effective chemical potential.
This procedure recovers a bulk density equal to the value of the density profile in the inhomogeneous system at position $x$.
As we show, LDA is a viable approximation for the fluctuation profiles as well.

In a very dilute Lennard-Jones gas ($\rho_b \approx 0.025\,\sigma^{-3}$) with $\alpha=0.2\,\varepsilon$ and $L=25\,\sigma$ shown in \cref{fig:LDA}(b), values of bulk systems coincide with the local values of the fluctuation profiles in the perturbed system.
This can also be verified for a Lennard-Jones liquid ($\rho_b \approx 0.78\,\sigma^{-3}$) in \cref{fig:LDA}(a) and for a dilute fluid of Gaussian particles ($\rho_b \approx 0.14\,\sigma^{-3}$) in \cref{fig:LDA}(c).

\end{document}